\documentclass[10pt]{article}

\usepackage[margin=1in]{geometry}
\usepackage[utf8]{inputenc}
\usepackage[english]{babel}

\usepackage{graphicx}
\usepackage{booktabs}
\usepackage{longtable}
\usepackage{amsmath,amssymb}
\usepackage{array,multirow}
\usepackage{xcolor}
\usepackage[T1]{fontenc}

\usepackage[numbers,sort&compress]{natbib}

\usepackage[colorlinks=true,linkcolor=blue,citecolor=blue,urlcolor=blue]{hyperref}

\usepackage{lineno}
\title{Mapping the influence of symmetry breaking in structure-property relationships of ABO$_3$ perovskites}

\author{
Panupol Untarabut$^{1}$, 
Sylvian Cadars$^{1}$, 
Fabien Pascale$^{2}$, \\
S\'ebastien Leb\`egue$^{2}$, 
Olivier Masson$^{1}$
Samuel Bernard$^{1}$\\
Assil Bouzid$^{1,\ast}$, 
Santanu Saha$^{1,\ast}$
}

\date{}

\begin{document}
\maketitle

% --- Affiliations ---
\noindent
$^{1}$ University of Limoges, CNRS, IRCER, UMR 7315, F-87000, Limoges, France\\
%%$^{1}$ Institut de Recherche sur les Céramiques (IRCER), UMR CNRS 7315, Universit\'e de Limoges, 87068 Limoges, France\\
$^{2}$ Universit\'e de Lorraine, CNRS, LEMTA, F-54000 Nancy, France\\[4pt]

%Université de Lorraine, LPCT, CNRS UMR7019, F-54506 Vandoeuvre-lès-Nancy, France\\[4pt]

% --- Corresponding authors ---
\noindent
$^{\ast}$Correspondence: assil.bouzid@cnrs.fr; santanu.saha@unilim.fr

\begin{abstract}
%%199 Words - Abstract Limit to 200 words for Nature Communication
Perovskite oxides have emerged as an important class of material with promising energy applications owing to their compositional and structural flexibility, which enables stabilization of both low- and high-symmetry phases and gives rise to diverse physical properties. Under ambient conditions, most perovskites adopt low-symmetry structures characterized by octahedral tilting and B-site displacements. Despite their importance, computational studies have largely focused on the ideal cubic phase as modeling these distortions remains challenging. The difficulty stems from the absence of a quantitative framework capable of capturing composition-dependent distortions that can occur through multiple non-equivalent atomic displacement modes, often requiring computationally expensive large supercells to explore the structural landscape. Consequently, the influence of distortions on the stability and properties of low-symmetry perovskites remains insufficiently understood. In this work, we develop an efficient computational framework for the rapid construction and exploration of composition-dependent structural models across both low- and high-symmetry phases. Using \textit{Symmetry-Constrained Templates} and \textit{unconstrained supercell templates}, we systematically investigate 15 representative compositions to uncover relationships between composition, supercell size and shape, and distortion patterns. Based on these insights, we propose a robust and computationally inexpensive protocol for rapid structural exploration and assess the influence of different distortion modes on key physical properties.
\end{abstract}


%\maketitle

\section*{Introduction}\label{sec:INTRODUCTION}

The material class of ternary perovskite oxides had been a fascinating playground for a wide range of research activities over the past century. Since its discovery in 1850s, activities until $\sim$2000 primarily focused on understanding structural phases, nature of distortions and their evolution due to external factors, primarily via experimental characterizations. With the advent of powerful \textit{ab initio} methodologies and precise characterization techniques, the focus gradually shifted towards understanding the atomistic role of chemical species on the origin of different emerging properties. The rising field of nanoscience further propelled developments of synthesis approaches in tailoring composition, shape and size for enhancing  efficiency in different energy applications~\cite{zhu_nanosized_2023,wu_recent_2023,xu_advancements_2024,seitz2016highly,suntivich2011perovskite,beall2021perovskite,hu2021identifying,xu2021perovskite,chu2022high,hwang2017perovskites,acosta2017batio3}. At the core, this evolving broadened interest of perovskites can be associated to (i) the implicit robustness of the ABO$_3$ lattice in hosting different chemical compositions  tolerant to substantial deformations~\cite{filip2018geometric,bartel2019new,emery2017high,bechtel2018octahedral,zhao2020polymorphous,feng2025unraveling,hwang2017perovskites,matheu2022halogen} and (ii) the synergistic influence of chemical species and structural distortion on their emerging properties~\cite{xu2025improper,yoshida2018ferroelectric,varignon2019origin}.

An apparently simple yet challenging step towards understanding this dynamically evolving field of perovskites has been the development of experimentally consistent structural models in accordance to structural disorder and chemical ordering in multinary compositions across the lattice.
Within an ideal picture, ABO$_3$ stabilizes in cubic phase (highest symmetry) where the larger A cation is
coordinated by 12 oxygen anions in a cubo-octahedral geometry (AO$_{12}$) while each B-cation is surrounded 
by 6 oxygen atoms, creating BO$_6$ octahedral units and establishing a rigid 3D framework. With introduction of asymmetry along certain lattice direction, the resulting structure can still retain high symmetry. But, with more complex set of asymmetries the structure gradually adapts a low symmetry phase.

One of the earliest attempts in understanding structural models of low symmetry perovskite phases had been by Goldschmidt in 1920's via the empirical tolerance factor ($t_G$)~\cite{goldschmidt1926gesetze,shannon1976revised}.
Later, this complex picture was greatly simplified by early works of Glazer in 1970's~\cite{glazer1972classification} and later by Stokes and Howard in $\sim$2000~\cite{howard1998group,stokes2002group}. Based on their crystallographic and mathematical reasoning, these structural distortions can be represented as a combination of different coherent in-phase and out-phase octahedral tilting distortions of  BO$_6$ octahedra. In practice, this boils down infinite possibilities to just 15 possible phases. Beyond tilting distortion, B-site atomic displacement within the BO$_6$ octahedra is still relevant for many interesting phenomena such as ferro-electricity and emerging quantum phases
~\cite{yoshida2018ferroelectric,varignon2019origin,tang2025regulating,seitz2016highly,suntivich2011perovskite,beall2021perovskite}. These simplified classification and nomenclature schemes proved valuable in identifying new phases of different perovskite compositions.

Over the decades, \textit{ab initio} methods coupled with other methodologies~\cite{oganov2011modern,woodley2008crystal,rapaport2004art,kuhne2014second} have become indispensable in identifying and correlating new-phases with structure-property relationships. While high symmetry phases can be easily explored with small unit cells, low symmetry phases required sampling over multiple structural configurations to obtain a representative average structural model. Two popular approaches have emerged for these explorations : (a) Soft phonon approach - Based on the calculated dynamical matrices and imaginary frequencies, this approach systemically tracks direction of atomic movements to apply suitable atomic distortions~\cite{vali2008structural,axe1969zone,ackland1997soft,pascale2022strategies} and (b) Supercell method - This involves construction of large supercells, introduction of random atomic perturbations and relaxing the structures using appropriate \textit{ab initio} methods\cite{zhao2020polymorphous}. While the former is a preferred choice for exploring high symmetry phases, later one is the default approach for low symmetry phases. Both methods are implicitly computationally intensive and the complexity can easily amplify in chemically disordered systems containing multiple A-site or B-site species.

On contrary to these two computationally expensive approaches, various broken-symmetry structural templates identified in Ref.\cite{howard1998group,glazer1972classification} could serve as an inexpensive route for rapid exploration. But, as the nature of distortions are strongly composition dependent with no well-defined trends, their wide scale adaptation becomes difficult in practice. In addition, these structural models have not been systematically validated across different compositions to highlight their success rate or their limitations against different approaches. Thus, while well known high-symmetry phases such as tetragonal, orthorhombic and hexagonal are well understood, exploration of various low-symmetry phases remains challenging and often limited to few selected composition of interest~\cite{el2011modeling,tan2021critical,xie2020generation}.

To the best of our knowledge, literature on perovskite oxides still remains largely skewed towards well reported high symmetry phases, whereas, extensive reports on low symmetry phases remain non-existent. A similar picture is illustrated by various open \textit{ab initio} materials databases. 
But, recent works on perovskite energy materials highlights the necessity to move beyond cubic phase to correctly capture local distortions for addressing real time stability and efficiency challenges under ambient conditions~\cite{tang2025regulating,seitz2016highly,suntivich2011perovskite,beall2021perovskite}. This rejuvenated interest has led to the initiation of modeling activities around the impact of different types of structural distortions on material properties~\cite{feng2025unraveling,bechtel2018octahedral,zhao2020polymorphous,yang2020assessment}.

Recently, multiple flavours of universal/foundational machine learning interatomic potentials (uMLIP) \cite{yang2024mattersim,batatia2022mace,batatia2025foundation,mazitov2025pet} have been released. These uMLIP in principle can serve as an accelerator for rapid exploration of low symmetry phases with large realistic supercell models. But, their transferability and accuracy across different simple/multi-elemental compositions is still prohibitive due to the limited diversity of perovskite databases as explained above. With no available datasets on low symmetry phases, it is simply not feasible to develop and train high quality MLIPs for diverse perovskite compositions. And, as explained above producing these set of \textit{ab initio} data poses multiple challenges.

Hence, based on the current situation, an in-depth understanding of physical and chemical properties of low-symmetry perovskite phases remains limited. Existing \textit{ab initio} based strategies are computationally expensive for rapid exploration. Various promising theorized structural templates (Ref.~\cite{howard1998group,glazer1972classification}) cannot be adapted efficiently within workflows due to unknown trends on distortion patterns, and the application of current uMLIPs is still limited due to transferability issues. These limitations pose a major bottleneck in rapid global exploration of both compositionally simple and complex perovskite oxides for various applications.

Our overarching goal has been on the development of a robust high-throughput workflow which enables rapid exploration of compositionally diverse perovskite oxides across both high symmetry and low symmetry phases. A critical starting point of such developmental work is the identification of relevant structural templates adherent to the physico-chemical nature of perovskite oxides. In this manuscript we address this challenge by first mapping symmetry lowering mechanisms and their relative importance on different physical properties. To achieve this, we carry out systematic \textit{ab initio} exploration of both high and low symmetry phases of 15 representative perovskite oxides, their thermodynamic ranking and emerging physical properties. Based on the cumulative understanding from this extensive study, we then present a simplistic adaptable robust framework for rapid construction of composition dependent high symmetry/low symmetry structural templates which enables efficient exploration.

The paper is organised as follows in different segments in Sec.\ref{sec1:RESU}: 
(a) We first present the set of criteria and the list of 15 representative perovskite compositions, and then 
(b) provide a detailed description on the set of \textit{Symmetry-Constrained Templates} and \textit{Unconstrained Templates}. 
In the next subsection (c) we present our analysis of thermodynamic ranking across different structural templates and analyse in depth the correlation between different octahedral distortion \textit{structural descriptors} and the thermodynamic ranking. We append this section with discussion on the influence of temperature on thermodynamic ranking of a selected set of symmetry-constrained templates
(d) In sub-section.~\textit{Physical Properties}, we extend the analysis of the influence of different distortions on key physical properties, particularly band gap and bulk modulus. (e) In sub-section.~\textit{Optimisation Protocol and Validation}, we discuss our new robust protocol for efficient exploration of low symmetry phases. Finally, in Sec. \textit{Conclusion} we present the main conclusions and outlook of this work. Details regarding computational setup and structural descriptors used to quantify octahedral tilting and B-site displacements are provided in Sec. \textit{Methods}.

\refstepcounter{section}
\section*{Results and Discussion}\label{sec1:RESU}

\subsection*{Selected Composition}
\phantomsection
\label{subsec1:COMPSELECT}
In order to explore and quantify the influence of structural distortions on thermodynamic properties, we constructed a set of representative compositions constrained to three key factors: 

(a) Absence of magnetism. To simplify exploration, we restricted compositions with B-site elements which do not display any magnetic ordering. For paramagnetic cases, we carried out non-magnetic calculations. 

(b) Chemical diversity. We picked $d$-block ($d^0$,$d^1$) and $p$-block elements for B-site, which typically govern the overall electronic behavior.

(c) Structural diversity. Qualitatively, it is well understood that the ionic radius of A-site cation controls the degree of octahedral tilting within perovskites and hence structural distortion. Using this, we explored suitable set of A-site cations. In addition, we adopted the Goldschmidt tolerance factor $t_G$ as an quantitative metric of structural distortion. The mathematical expression of $t_G$ is shown in eq.~(\ref{eq2:goldschmidt}) where $r_A$, $r_B$ and $r_O$ represent Shannon ionic radii of A-site, B-site cation and O anion, respectively.~\cite{goldschmidt1926gesetze,shannon1976revised}.

\begin{equation}\label{eq2:goldschmidt}
t_G = \frac{r_A + r_O}{\sqrt{2}(r_B + r_O)}
\end{equation}

\begin{figure}[!thb]
\centering
\includegraphics[scale=0.90]{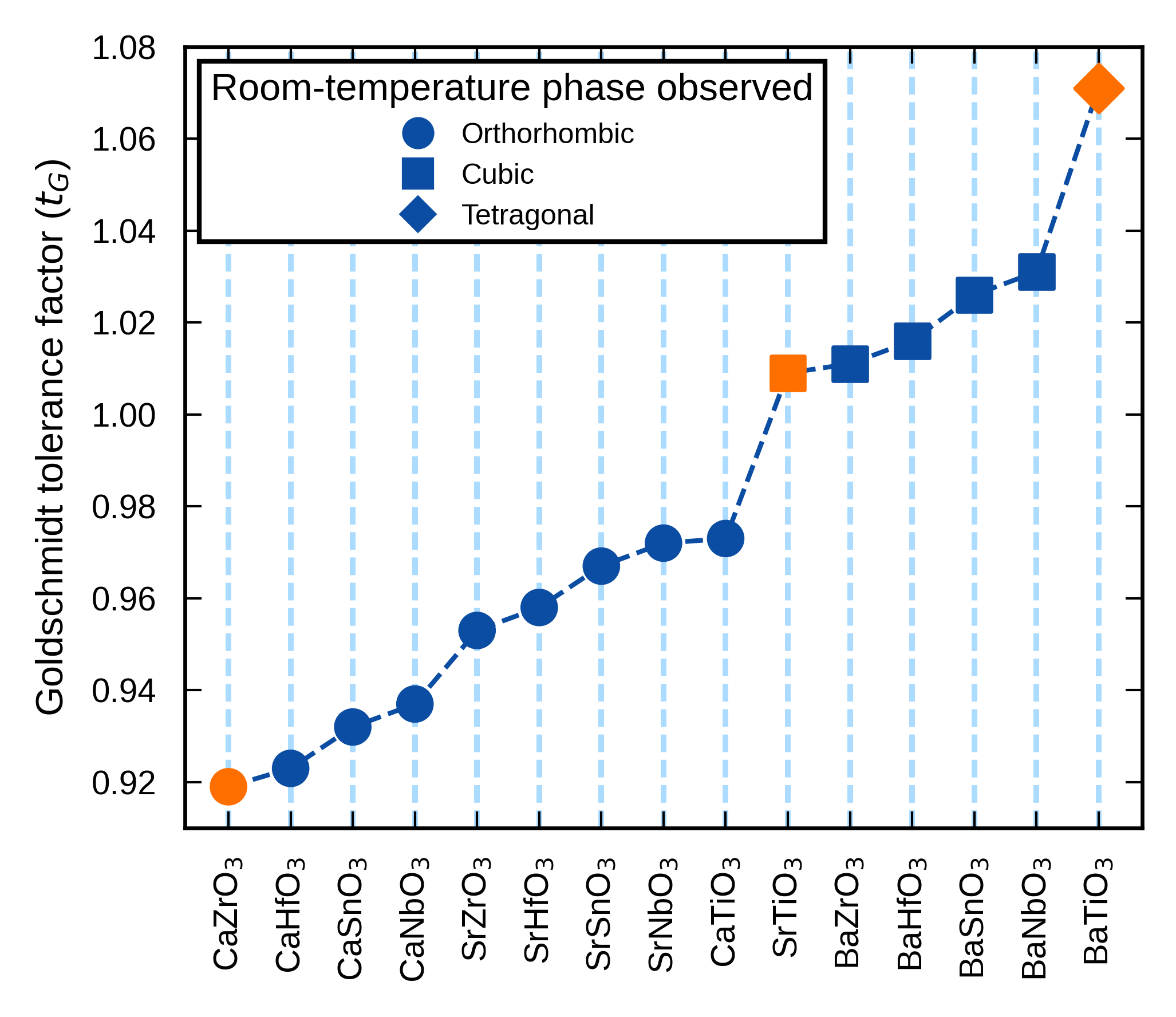}
\caption{Variation of Goldschmidt tolerance factor $t_{\text{G}}$ (y-axis) across a series of ABO$_3$ perovskite compounds (x-axis) arranged in ascending order. The experimentally reported phases at room temperature are represented by different markers: circles (orthorhombic), squares (cubic), and diamonds (tetragonal). Orange markers denote the hand-picked compositions for detailed \textit{ab initio} studies; CaZrO$_3$, SrTiO$_3$, and BaTiO$_3$.}
\label{fig01:TolFacG}
\end{figure}

Adopting these constraints, the list boils down to 15 representative compositions formed by the combination of A = [Ca, Sr, Ba] and B = [Ti, Zr, Hf, Sn, Nb] elements. While their chemical diversity can be interpreted from the valence states (orbitals/occupations) of B-site elements in +4 oxidation state, their structural diversity is illustrated by the computed $t_G$ values (0.90 $< t_G <$ 1.08) as shown in Fig.~\ref{fig01:TolFacG}. Taking into account the experimentally reported perovskite phases of these compositions under ambient conditions (markers in Fig.~\ref{fig01:TolFacG}), $t_G$ values correlates well with their propensity to stabilise in a low-symmetry phase ($t_G < 1.00$), the ideal cubic phase ($t_G \approx 1$), or a non-perovskite phase ($t_G > 1.00$). Based on this validation, we handpick CaZrO$_3$ ($t_G = 0.919$), SrTiO$_3$ ($t_G = 1.009$), and BaTiO$_3$ ($t_G = 1.071$) as prototypical examples for detailed investigation as marked by orange markers. Systematic tests with other reported tolerance factors~\cite{bartel2019new} led to similar trends. Additional details are provided in Table.~S2 in Sec.~S1 of the SI.

%%%%%%%%%%%%%%%%%%%%%%%%%%%%%%%%%%%%%%%%%%%%%%%%%%%%%%%%%%%%%%%%%%%%%%%%%%%%%%%%%%%%%%%%%%%%%%%%%%%%%%%%%%%%%%%%%%%%%%%%%%%%%%%%%%%%%%%%%%%%%%%%%%%%%%%%%%%

\subsection*{Templates Description}
\label{subsec2:STRUCTEMPLATES}
%%\section*{Results and Discussion}\label{sec:RESU}\saha{CHECK}

Among different possible distortions, ABO$_3$ perovskite oxides are known to be strongly influenced by  
the tilting of BO$_6$ octahedra ~\cite{glazer1972classification,howard1998group,simpson2025octahedral} and the B-site cation displacements.~\cite{cohen1990lattice,cohen1992origin,yoshida2018ferroelectric,xu2025improper}. Adopting appropriate structural models embodying these set of complicated distortions is essential for systematic sampling of their thermodynamic landscape. 

In this work, we therefore explore two kinds of structural templates: (i) \textit{Symmetry-constrained templates} which embody these two well-known types of distortions via \textit{Glazer Templates} and \textit{Standard Templates}   
and (ii) \textit{Unconstrained Templates} via \textit{Supercell Templates} of different shapes, sizes and parity (as will be explained below)~\cite{zhao2020polymorphous}  and \textit{Distorted Templates}~\cite{dias2021role} where the probed local distortions extend beyond predefined patterns.

\subsubsection*{Symmetry Constrained Templates}

%%\subsubsubsection*{Glazer Templates}
\textbf{Glazer Templates}\\ 
In this section, we first focus on how different combinations of cooperative BO$_6$ octahedral rotations along one, two, and three crystallographic directions affect the thermodynamic stability of ABO$_3$ perovskites.

\begin{figure}[!htb]
\centering
\includegraphics[scale=0.90]{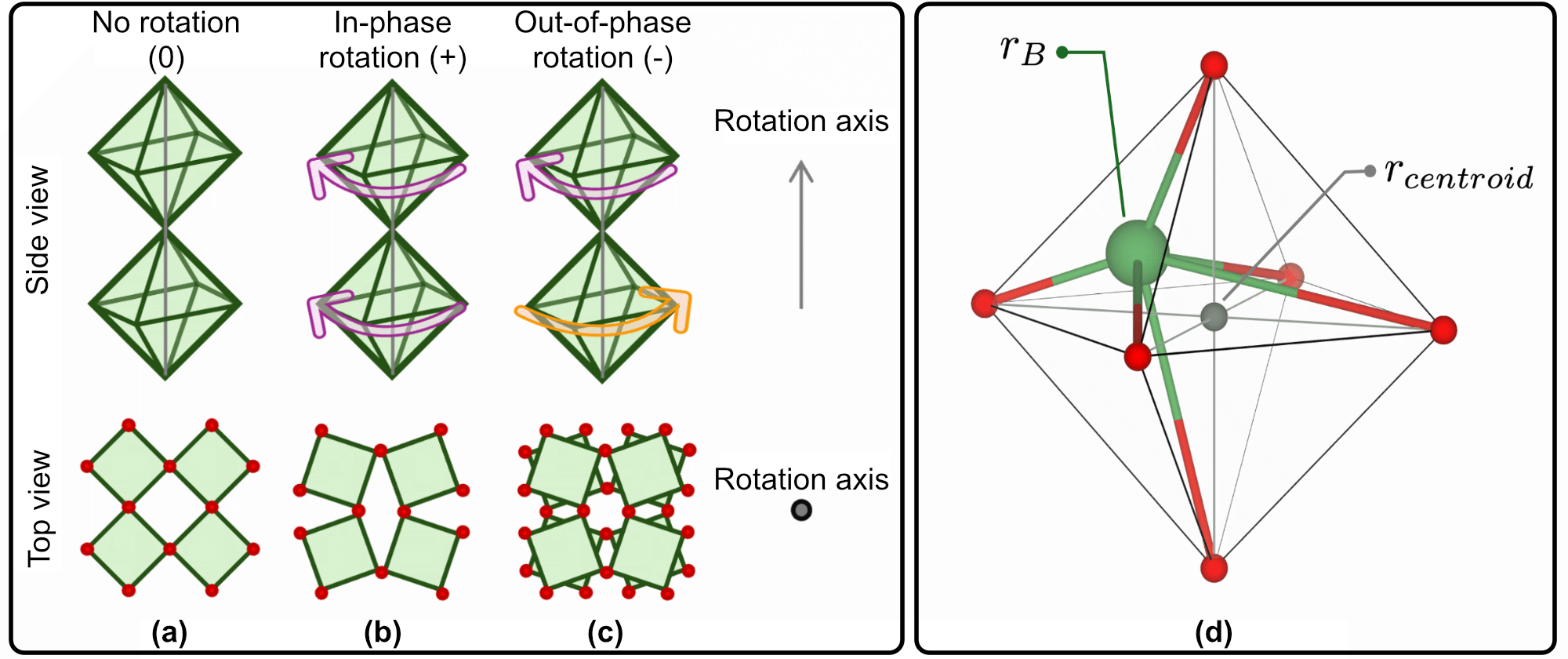}
\caption{\label{fig03:Octa-Rot}
Schematic illustration of BO$_6$ octahedral rotation modes in perovskite structures. 
(a) Ideal cubic structure with no octahedral rotation. 
(b) In-phase rotation (IR), where two adjacent octahedron rotate in the same direction along the rotation axis. 
(c) Out-of-phase rotation (OR), where two adjacent octahedron rotate in opposite directions. Side views (top row) highlight the relative orientation along the rotation axis, while top views (bottom row) show the projection of rotation patterns in the plane perpendicular to the rotation axis. 
(d) B-site off-centre displacement within a BO$_6$ octahedron. Green and red spheres represent B and O atoms, respectively}
\end{figure}

To construct structural models with well-defined octahedral rotation patterns, we adopt the classification scheme developed by Glazer~\cite{glazer1972classification}. Following well-defined terminology on Glazer notation, the three letters $a$, $b$, and $c$ correspond to the three crystallographic directions. Identical letters indicate that the magnitude of the rotation is same along the corresponding directions, while different letters indicate different magnitudes. The superscripts $0$, $+$, and $-$ denote, respectively, no rotation, in-phase rotation (IR), and out-of-phase rotation (OR) as illustrated in Fig.~\ref{fig03:Octa-Rot} about each axis. For example, $a^0a^0c^+$ describes IR only about the $c$ direction, whereas $a^+b^-b^-$ describes one direction with IR and two other directions with OR of equal magnitude. In practice, we generate these patterns by rotating neighbouring BO$_6$ octahedra in the same direction for a $+$ label and in alternating directions (clockwise/anticlockwise) for a $-$ label along the chosen axis. At the initial step, the A and B-site cations are kept fixed at their ideal positions and only the oxygen atoms are rotated. This exploration of tilt-systems is limited to the 15 unique tilt-systems compatible with perovskite framework (1 no-tilt, 2 one-tilt, 4 two-tilt, and 8 three-tilt systems), as identified by Stokes and Howard~\cite{howard1998group}. We refer to this set of structural models as the \textit{Glazer Templates}. Details pertaining to the Glazer symbols, associated space groups, size of unit cells and a detailed description of the construction are summarized in Table~S3 (Sec.~S4(A1) of the SI). \\

%%\subsubsubsection*{Standard Templates}
\textbf{Standard Templates} \\
Most of the published literature and material databases on perovskite oxides primarily focus on well reported phases such as the cubic, tetragonal, orthorhombic and hexagonal phases. And, many of them can be categorised as one of the \textit{Glazer Templates}. But, the outliers can be marked by the presence of B-site octahedral distortion and other distortions going beyond the $BO_6$ octahedral tilting.

To construct a minimal map of the influence of different kind of distortions within symmetry constrained templates, we group five reference templates as representative of these well reported phases: cubic ($Pm\overline{3}m$, 221), tetragonal-I ($P4mm$, 99), tetragonal-II ($P4/mbm$, 127)~\cite{yoshida2018ferroelectric,stokes2002group}, orthorhombic ($Pnma$, 62)~\cite{YASHIMA2009120,knight2011structural} and hexagonal ($P6_3/mmc$, 194). Among them, cubic, tetragonal-II and orthorhombic templates are equivalent to $a^0a^0a^0$, $a^0a^0c^+$ and $a^+b^-b^-$ \textit{Glazer Templates}, respectively. The tetragonal-I template serves as an representative of a template with B-site displacement distortion (ignored within \textit{Glazer Templates}). Finally, the hexagonal template acts as an outlier featuring non-perovskite structure with face and corner sharing octahedra~\cite{sinclair1999structure,nenert2007experimental}. We refer to this collection of templates as \textit{standard templates} and the structural models are shown in Fig.~\ref{fig02:Standard-Templates}(a).

\subsubsection*{Unconstrained Templates}

%%\subsubsection*{Supercell Templates}
\textbf{Supercell Templates}\\ 
We now introduce \textit{Supercell Templates}, i.e. supercells based on the primitive cubic unit cell of ABO$_3$ that incorporate initial random displacement on the atomic sites and wherein relaxation is not constrained by symmetry. The objectives of this approach are three fold: (i) sample a broader portion of the potential energy surface, including distortion patterns that may not be captured by a finite catalogue of \textit{Symmetry-Constrained Templates}, (ii) assess how supercell shape, size, and parity (even vs.\ odd multiples) influence the emergence of different structural distortions (octahedral tilting, B-site distortion) and (iii) evaluate the influence of local disorder on relative thermodynamic stability~\cite{xie_generation_2020,Lee_2016,bechtel2018octahedral,zhao2020polymorphous}.

\begin{figure*}[!thb]
\centering
\includegraphics[scale=0.90]{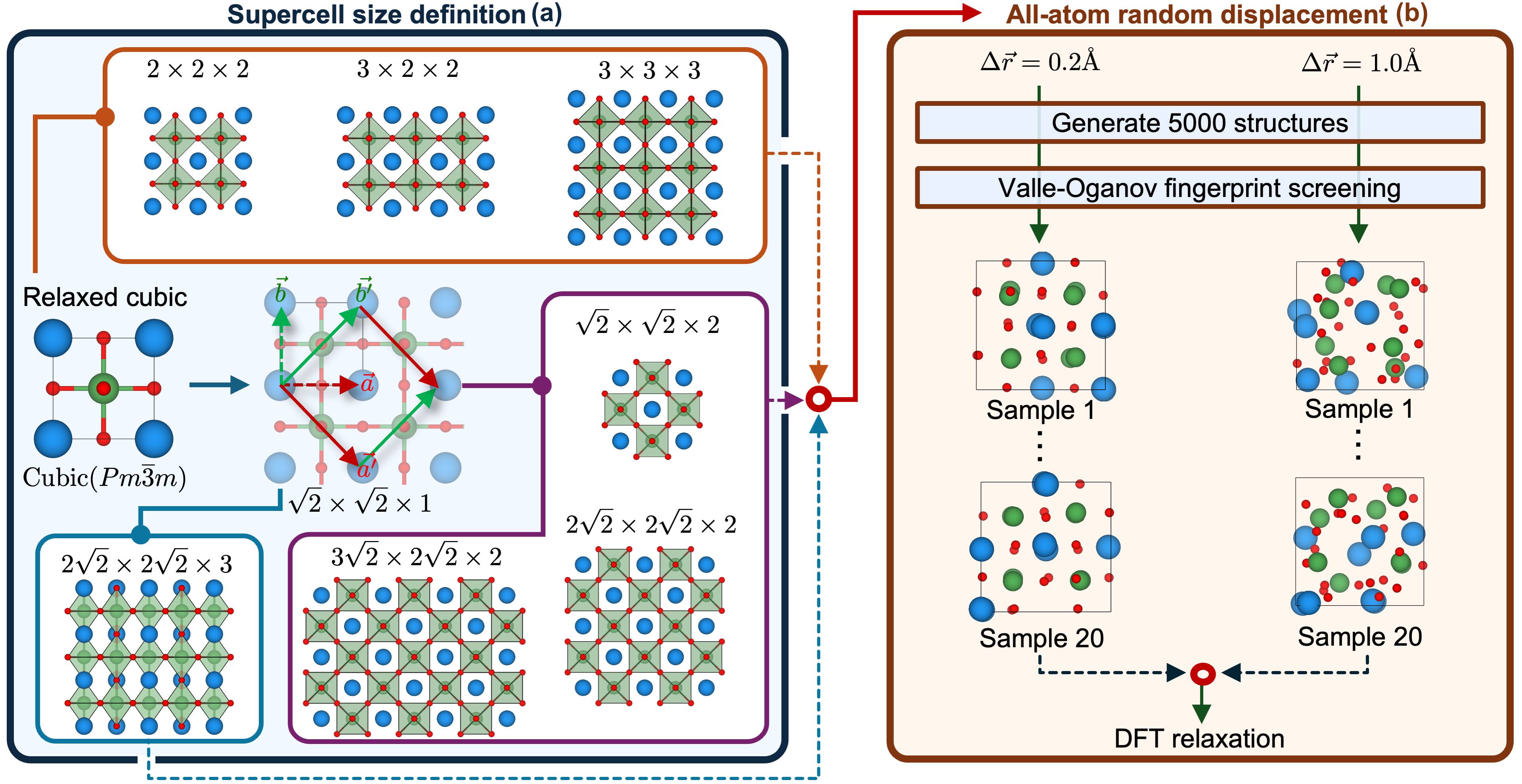}
\caption{\label{fig05:Flowchart-Supercell}Schematic workflow for generating distorted supercell structures for DFT calculations.(a) Displays different supercell types and their expansion and (b) Represents protocol followed to generate supercells with different random perturbations}
\end{figure*}

As discussed above %%in Sec.\ref{subsec1:COMPSELECT}, 
the \textit{Supercell Templates} are explored only on three representative compounds: CaZrO$_3$, SrTiO$_3$ and BaTiO$_3$ representative of low, intermediate and high $t_G$ cases, respectively.
To generate  initial random structural models in a systematic manner, we followed the workflow as summarised in Fig.~\ref{fig05:Flowchart-Supercell}. Starting from the relaxed cubic $Pm\overline{3}m$ structure, we construct a set of supercells using different expansion matrices. These supercells can be grouped into three families:

\underline{(i) \textit{Simple integer expansions}} of the primitive cubic cell along the three lattice vectors: \ $2 \times 2 \times 2$, $3 \times 2 \times 2$, and $3 \times 3 \times 3$. Here, $2 \times 2 \times 2$ serves as minimal representation of only even expansion, $3 \times 3 \times 3$ as a minimal representation of only odd expansion. The expansion $3 \times 2 \times 2$ represents a mixed case with odd expansion in one lattice direction only.

\underline{(ii) \textit{Rotated-lattice expansions with even parity}} based on a pseudocubic $\sqrt{2} \times \sqrt{2} \times 1$ cell, obtained by applying the $3 \times 3$ transformation matrix in eq.(\ref{eq:matrix}):

\begin{equation}\label{eq:matrix}
\begin{pmatrix}
1 & 1 & 0 \\
-1 & 1 & 0 \\
0 & 0 & 1
\end{pmatrix}
\end{equation}

which yields lattice vector lengths $(a, b, c) = (\sqrt{2}, \sqrt{2}, 1)$ in reduced cubic units. Within this representation, the $a$ and $b$ directions contain chains of two corner-sharing BO$_6$ octahedra, whereas the $c$ direction contains a single BO$_6$ unit. For brevity, we denote this cell as $\sqrt{2} \times \sqrt{2} \times 1$ and its diagonal expansions as $n\sqrt{2} \times l\sqrt{2} \times m$, i.e.\ $\sqrt{2} \times \sqrt{2} \times 2$, $2\sqrt{2} \times 2\sqrt{2} \times 2$, $3\sqrt{2} \times 2\sqrt{2} \times 2$, and $2\sqrt{2} \times 2\sqrt{2} \times 4$. The largest rotated supercell, $2\sqrt{2} \times 2\sqrt{2} \times 4$ leads to same number of formula units as reported in the recent works of Zunger and co-workers on polymorphous ternary perovskites.~\cite{zhao2020polymorphous}. Across all these constructed supercells, the number of corner-sharing BO$_6$ units always has even parity along all three crystallographic directions.

\underline{(iii) \textit{Rotated-lattice expansions with odd parity}} along the $c$ direction, $2\sqrt{2} \times 2\sqrt{2} \times 3$. In this case, the $a$ and $b$ directions still contain four BO$_6$ units, but the $c$ direction contains three BO$_6$ units. This setup constraints the representation of an OR pattern along $c$ direction.

Hence, the three families of \textit{Supercell Templates} illustrated in Fig.~\ref{fig05:Flowchart-Supercell}(a) provide unique advantages in terms of representing or constraining different octahedral tilting patterns. Recent studies and our current work suggest that while the $\sqrt{2} \times \sqrt{2} \times 1$-based cells are particularly suited to accommodate common energy lowering low-symmetry tilt modes ($a^0a^0c^+$) observed in perovskites~\cite{Lee_2016, Jung_2025}, the integer expansions of the primitive cell provide a direct route to both simple and complex  
periodicities. In combination, they allow us to probe the effect of supercell parity along each direction.

For each supercell size, we generate 5000 randomly distorted configurations by applying atomic displacements of fixed amplitude $|\Delta \vec{r}| = 0.2$~\AA{} or 1.0~\AA{}, corresponding to low and high initial distortion levels, respectively (Fig.~\ref{fig05:Flowchart-Supercell}(b)). For each atom, the three cartesian components of the displacement vector are drawn from a uniform distribution in $[-1, 1]$, normalised to a unit vector, and then scaled by the target amplitude. From the resulting pool, we select the 20 most structurally distinct configurations per amplitude and supercell using the Valle–Oganov fingerprint metric~\cite{oganov2009quantify}; thus, maximising structural diversity while keeping the number of DFT relaxations tractable. 

Applying this workflow on each composition, and for all supercells but the $2\sqrt{2} \times 2\sqrt{2} \times 4$, we  generated 280 distinct initial configurations ($7$ supercell sizes $\times$ 2 amplitudes $\times$ 20 structures). For the largest rotated supercell, $2\sqrt{2} \times 2\sqrt{2} \times 4$, we reduced the number of selected structures to 10 per amplitude to limit the computational cost, yielding an additional 20 configurations. In total, each compound was thus sampled with 300 distinct initial structures across all supercell families. Further details on the structure generation and selection protocol are provided in Sec.~S4(B1) of the SI.

\textbf{Distorted Templates}\\
The above discussed structural templates result from multiple combinations of distortions and thereby makes it difficult to breakdown the relative importance of each type of distortion on improving thermodynamic stability. To quantify the relative importance of each distortion type, we present \textit{distorted templates} where we introduce only one kind of distortion in the structure at a time. Here, we particularly focus on three generic classes of distortions: (i) in-phase octahedral rotation (Fig.\ref{fig03:Octa-Rot}(b)), (ii) out-of-phase octahedral rotation and Fig.~\ref{fig03:Octa-Rot}(c) (iii) displacement of the B-site cation within the BO$_6$ octahedron (Fig.~\ref{fig03:Octa-Rot}(d)).

Recently, Dias et al.~\cite{dias2021role} carried out detailed study on influence of these kind of distortions on halide perovskites via introduction of few representative structural models. We realize the above discussed distortion by adopting the following four structural templates in Ref.[\cite{dias2021role}]: (i) distorted-cubic (d-cubic) incorporating B-site displacement achieved by elongation/shortening of B-O bond in a primitive cubic cell, (ii) distorted-tetragonal with in-phase rotation (d-tetragonal IR) incorporating IR tilting only along $c$-direction in a $\sqrt{2} \times \sqrt{2} \times 2$ supercell, (iii) distorted-tetragonal with out of phase rotation (d-tetragonal OR) incorporating OR tilting only along $c$-direction in a $\sqrt{2} \times \sqrt{2} \times 2$ supercell. For consistency with ref ~\cite{dias2021role}, (iv) the super-cubic (s-cubic) template with cell expansion $2 \times 2 \times 2$ is also introduced and serves as an unbiased, symmetry-breaking starting point (similar by construction to the \textit{Supercell Templates} of corresponding size). In all \textit{Distorted Templates}, A-site/B-site cations are displaced randomly by a magnitude of 0.5~\AA. Additional details on the construction of these templates can be found in Sec.~S4(B2) of the SI.

%%%%%%%%%%%%%%%%%%%%%%%%%%%%%%%%%%%%%%%%%%%%%%%%%%%%%%%%%%%%%%%%%%%%%%%%%%%%%%%%%%%%%%%%%%%%%%%%%%%%%%%%%%%%%%%%%%%%%%%%%%%%%%%%%%%%%%%%%%%%%%%%%%%%%%%%%%%

\subsection*{Thermodynamic Ranking\label{sec3:THERMODYNAMIC}}
\label{subsec3:THERMODYNAMIC}
\subsubsection*{Symmetry-Constrained Templates\label{sub:THERMOGLAZER}}
%%\subsubsubsection*{\label{subsec:GLAZERTEMP}Glazer Templates}

\begin{figure*}[!htb]
\centering
\includegraphics[scale=0.90]{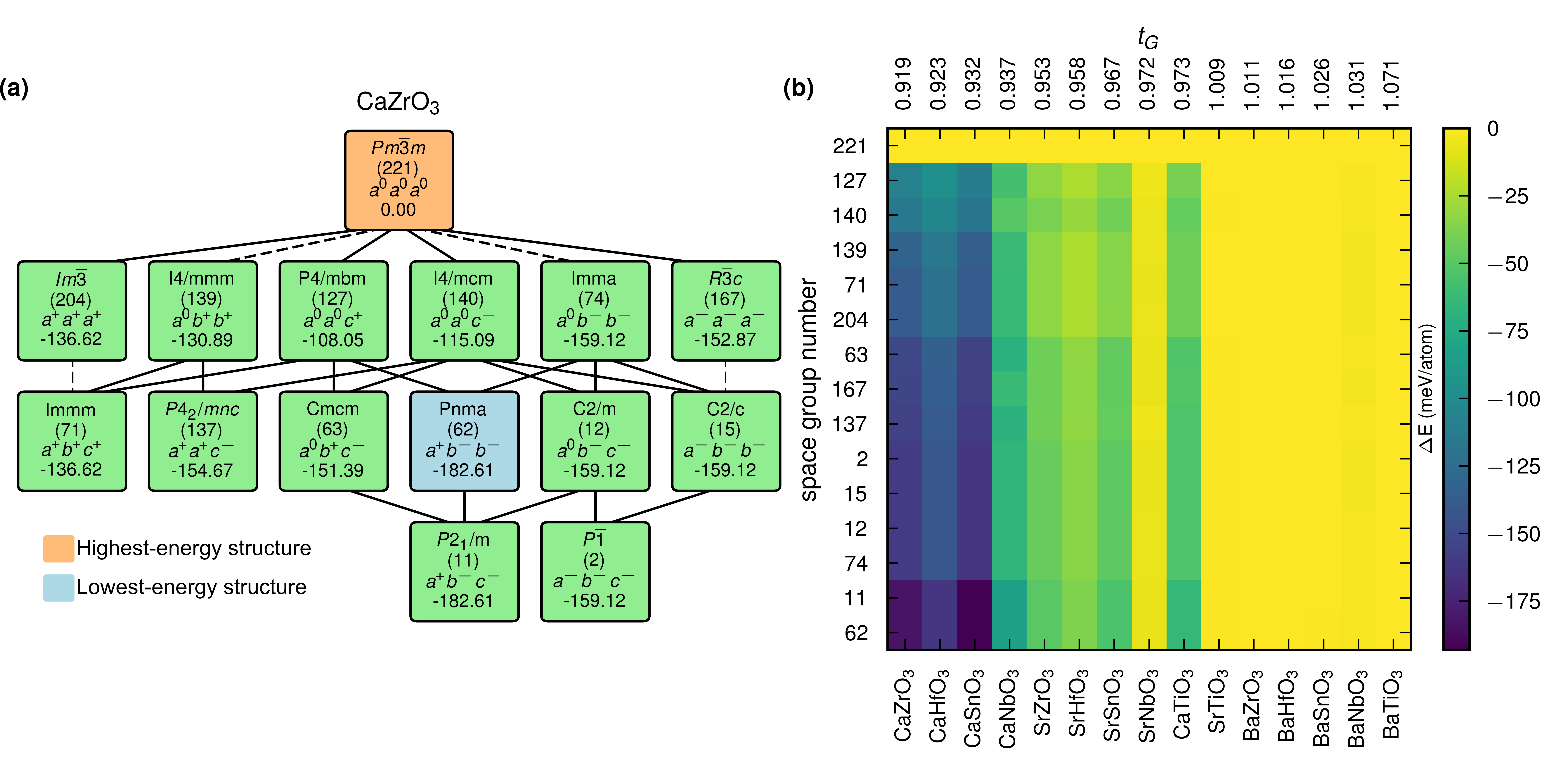}
\caption{\label{fig04:Glazer-Templates}(a) Group–subgroup relationships of Glazer templates for CaZrO$_3$. Each box shows the space group symbol, space group number, corresponding Glazer notation, and the relative energy (meV/atom). Solid lines between boxes indicate allowed continuous group–subgroup transitions, while dashed lines represent first-order transitions as classified by Landau theory. (b) Heatmap of relative energies $\Delta E$ (meV/atom) for all Glazer templates across the full set of 15 ABO$_3$ compositions. The y-axis corresponds to the space group number, organised according to the thermodynamic ranking for CaZrO$_3$, and the x-axis lists compositions ordered by increasing Goldschmidt tolerance factor $t_{\text{G}}$. Energies are given relative to the cubic phase, with darker colours indicating lower (more stable) configurations.}
\end{figure*}

\textbf{Glazer Templates} \\
Adopting the \textit{Glazer} templates with appropriate lattice parameters for all 15 compositions, we performed full geometry relaxations and analysed their thermodynamic ranking for each composition. As a representative example, we first discuss CaZrO$_3$ with the lowest $t_G$ (0.919) in our set, as shown in Fig.~\ref{fig04:Glazer-Templates}(a). Each box represents a different template, labelled by the space group symbol, space group number, Glazer notation, and relative energy $\Delta E$ with respect to the ideal cubic phase.

The orange box at the top, with Glazer notation $a^0a^0a^0$, represents the ideal cubic phase. One-tilt templates introducing octahedral tilting along a single direction, such as $a^0a^0c^-$ (140) and $a^0a^0c^+$ (127), already improve the stabilization energy by –115~meV/atom and –108~meV/atom, respectively. Two-tilt templates further improve the stabilization energy within the range –160~$< \Delta E <$~–130~meV/atom. Finally, most three-tilt templates stabilize within a similar energy window as the two-tilt templates, with the notable exception of $a^+b^-b^-$ (62) and $a^+b^-c^-$ (11), which yield $\Delta E = -183$~meV/atom. This additional stabilization of approximately $\sim$25 meV/atom can be linked to the specific combination of one IR and two OR tilting patterns embodied in the (degenerate) $Pnma$ (62) and P2$_1$/m (11) \textit{Glazer Templates}. This lowest-energy template, $a^+b^-b^-$ is highlighted with a blue box in Fig.~\ref{fig04:Glazer-Templates}(a).

It is also important to note that several low-symmetry CaZrO$_3$ \textit{Glazer Templates} converge after relaxation to higher-symmetry templates. These include the transformation of $a^+b^-c^-$ (11) to $a^+b^-b^-$ (62), $a^0b^-c^-$ (12) to $a^0b^-b^-$ (74), and $a^+b^+c^+$ (71) to $a^+a^+a^+$ (204), leading to degenerate values of $\Delta E$. On closer inspection, these transformations correspond to a collapse of the initial asymmetry in the magnitudes of the octahedral tilting angles such that $|b|$ and $|c|$ become equal, while the underlying tilt pattern is preserved. In other words, the relaxation drives the system towards a higher-symmetry point on the same tilt-manifold, where the in-equivalence between tilt magnitudes along different axes is no longer energetically favoured.

On closer inspection, we observe that the $a^+b^-c^-$ (11) template never remains as a distinct local minimum after relaxation for any of the 15 ABO$_3$ compositions: it systematically relaxes to $Pnma$ ($a^+b^-b^-$, 62). This suggests that, within this set of simple ternary perovskites at 0~$K$, the $a^+b^-c^-$ tilt pattern does not correspond to a robust metastable structure; instead, the energy minimum lies along the symmetry line where the OR tilts in the $b$ and $c$ directions adopt equal magnitudes. This observation does not preclude the stabilisation of $a^+b^-c^-$ type tilt systems in other perovskites (such as SrRuO$_3$~\cite{vailionis2011misfit}); rather, it indicates that, for the present ABO$_3$ series of compounds, $Pnma$ is the preferred endpoint of relaxation when starting from this lower symmetry structural model.

Extension of a similar analysis of the thermodynamic ranking of different \textit{Glazer Templates} to the full set of compositions is shown in the heatmap in Fig.~\ref{fig04:Glazer-Templates}(b). While the compositions are ordered by increasing $t_G$ along x-axis, their relative energy $\Delta E$ (meV/atom) of different templates are ordered by thermodynamic ranking as found for CaZrO$_3$ with respect to the cubic phase. The group–subgroup diagrams for all 15 compositions are shown in Fig.~S3-S17 (Sec.~S4(C)).

Two main regions can be identified. On the right-hand side of Fig.~\ref{fig04:Glazer-Templates}(b), for compositions with $t_G \geq 1.00$, the energies are close to zero ($|\Delta E| \leq 5$~meV/atom) and most templates are nearly degenerate with the cubic reference. On the left-hand side of Fig.~\ref{fig04:Glazer-Templates}(b), for compositions with $t_G < 1.00$ (with exception of SrNbO$_3$), many templates show much larger negative $\Delta E$. In such low-$t_G$ systems, adopting different rotation patterns leads to large differences in energy. This is consistent with the picture that, for $t_G < 1.00$, the A-site cation is relatively small compared with the size of its cavity in the ideal cubic structure. Cooperative rotations of the BO$_6$ network reduce the volume of this cavity, shorten the A–O bond distances, and hence, lower the total energy. Different rotation patterns change the cavity shape in different ways, so the total energy depends strongly on which tilt system is adopted. In contrast, for larger $t_G$ the cubic framework is already close to the preferred geometry for the A-site cation, and rotations only provide a small additional gain; so the energies of the different templates become very similar. As $t_G$ increases, the spread in $\Delta E$ among templates gradually decreases and eventually falls below $\sim$1 meV/atom. \\

%%\subsubsubsection*{\label{subsec:STANDARDTEMP}Standard Templates}

\begin{figure*}[!htb]
\centering
\includegraphics[scale=0.90]{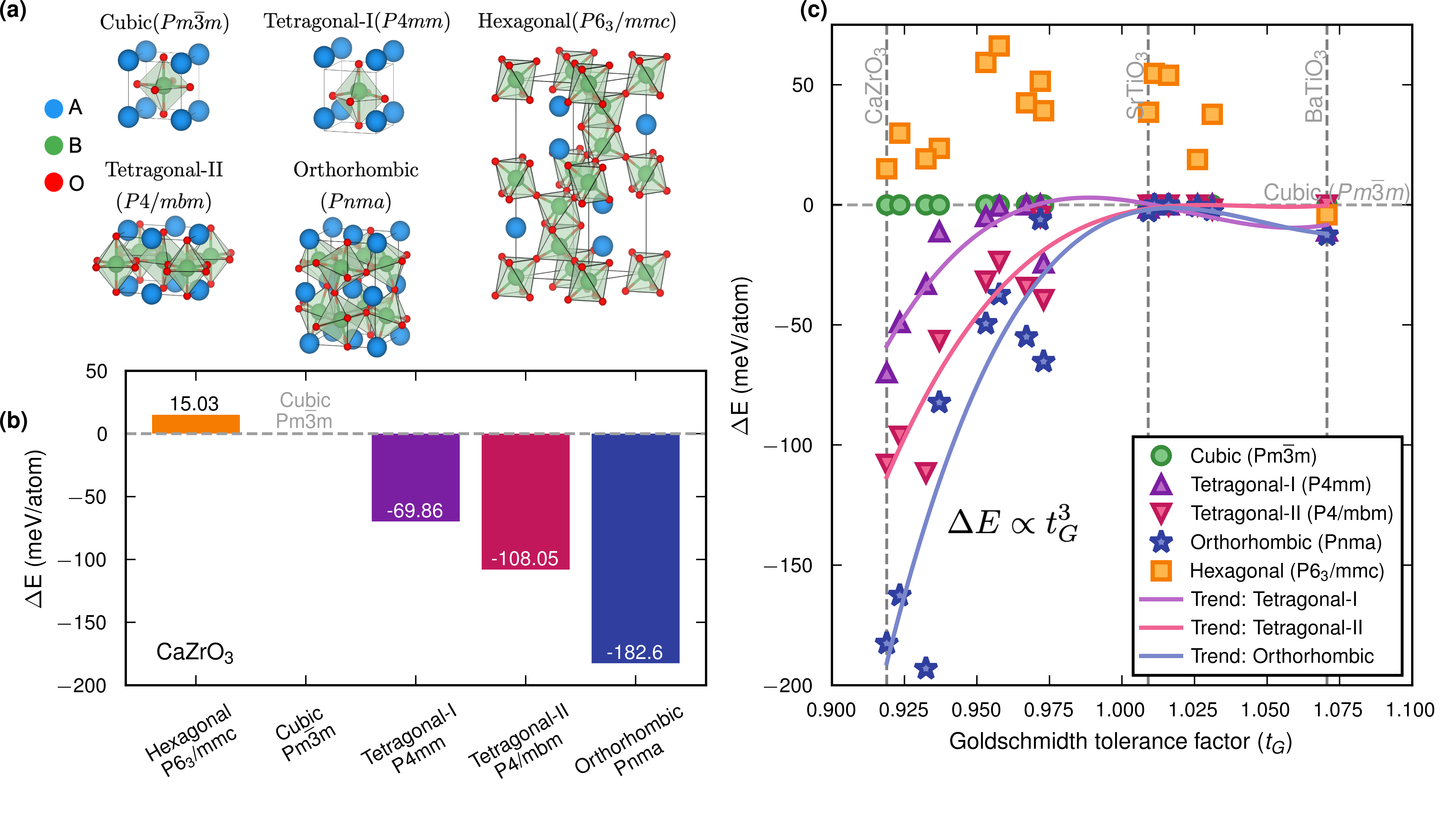}
\caption{\label{fig02:Standard-Templates}(a) Five distinct structural models of the ABO$_3$ in standard template. Blue, green, and red spheres represent A, B, and O atoms, respectively. BO$_6$ octahedra are represented by green polyhedra. (b) Relative energies of all five phases of CaZrO$_3$, with respect to the cubic phase ($Pm\overline{3}m$).  
(c) Relative energies of all 15 ABO$_3$ compositions as a function of $t_{\text{G}}$. The vertical dashed lines indicate the lowest (CaZrO$_3$; 0.919), highest (BaTiO$_3$; 1.071), and nearest-to-ideal (SrTiO$_3$; 1.009) tolerance factors. Each symbol represents a specific crystal template as shown in the legend. The solid lines represent a third-order polynomial fit of $\Delta E$ as a function of $t_G$ for orthorhombic (blue), tetragonal-II (magenta), and tetragonal-I (purple) templates, respectively}
\end{figure*}

\textbf{Standard Templates} \\
We now present the influence of octahedral tilting and other distortion types via the five \textit{standard templates} as shown in Fig.~\ref{fig02:Standard-Templates}(a). Following the discussion above on \textit{Glazer Templates}, we first present the results of this exploration on CaZrO$_3$ as shown in Fig.~\ref{fig02:Standard-Templates}(b). Among the considered templates, the orthorhombic ($Pnma$) phase turned out to be most stable, with an energy of $\sim$183 meV/atom below the cubic phase ($Pm\overline{3}m$). The tetragonal-II ($P4/mbm$) phase shows an intermediate thermodynamic ranking with -108 meV/atom with respect to the cubic phase. Their overall ranking is consistent with the results of the corresponding \textit{Glazer Templates} ($a^0a^0a^0$, $a^+b^-b^-$ and $a^0a^0c^+$). Among the outliers, the tetragonal-I ($P4mm$) phase with B-site distortion improves stability with –70 meV/atom, whereas the hexagonal ($P6_3/mmc$) template is slightly higher in energy than the cubic phase by $\sim$15 meV/atom. 

Assessment of the stability trend across all 15 compositions is shown in Fig.~\ref{fig02:Standard-Templates}(c). We observe that the orthorhombic template is consistently the most stable phase and the hexagonal template the least stable phase among all the standard templates. In general, all compositions display the following stability trend: Orthorhombic $>$ Tetragonal-II $>$ Tetragonal-I $>$ Cubic $>$ Hexagonal. Hence, the introduction of distortions through octahedral tilting and B-site displacements leads to lowering of energy and stabilization.~\cite{cohen1990lattice}.

Expectedly, the lowering of energy for each template with respect to the cubic phase appears to be strongly correlated with the $t_G$. This is clearly displayed by the fitted third-order polynomial trend lines to the relative energies of the orthorhombic (blue), tetragonal-II (magenta), and tetragonal-I (purple) templates as functions of $t_G$ in Fig.~\ref{fig02:Standard-Templates}(c) (eq.(S7),Sec.~S4(A2) of SI). These results clearly demonstrate that as $t_G$ increases, the energy differences among different structural phases systematically decrease, indicating that higher-symmetry phases are favored in compositions where the ratio of A-O and B-O bond distance is close to $\sim\sqrt{2}$.
%the ionic radii ratio between A- and B-site cations is highest. 
For compositions with $t_G \geq 1.00$, the energy differences between cubic and other high-symmetry phases are small ($\Delta E < 15$~meV/atom), consistent with the fact that these systems are often reported as cubic under ambient  conditions~\cite{de1996high,akbarzadeh2005combined,xin2019single,chahib2023investigation,vidya2013optical}. However, an interesting exception among these compositions is BaTiO$_3$, for which the hexagonal phase is predicted to be slightly more stable (–4.2 meV/atom)  
than the cubic phase, suggesting that at higher $t_G$ values, alternative non-perovskite structures may achieve comparable or greater stability as compared to classical perovskite phases~\cite{tenne2004absence,sinclair1999structure,glaister1960investigation}.

These results show that even a small set of \textit{standard templates} already captures the main effect of symmetry breaking on phase stability across the 15 compositions. However, these templates still represent only a limited subset of the possible distortion patterns of the BO$_6$ network and B-site positions.

It is noteworthy to point that despite this strong dependence of $\Delta E$ on $t_G$ for \textit{Symmetry-Constrained Templates}, all compositions retain the orthorhombic  $Pnma$ \textit{Standard template} (or equivalently the $a^+b^-b^-$ \textit{Glazer template}) as the lowest-energy structure, while the cubic $Pm\overline{3}m$ (or equivalently $a^0a^0a^0$ \textit{Glazer template}) remains the highest. The consistent emergence of the orthorhombic $Pnma$ template as the most stable tilt system across all compositions highlights its fundamental role and its prominent stability within the thermodynamic landscape of perovskite oxides. This trend is also in line with experimental observations that many perovskites which are cubic at ambient or high temperature transform to orthorhombic $Pnma$ phase upon cooling, as validated with the recent theoretical studies of temperature-dependent phase stability in perovskites (including halide analogues)~\cite{zhao2020polymorphous}. Thus, the \textit{Symmetry Constrained Templates} provide a simple and systematic way to quantify how the relative energies of different rotation patterns alter with $t_G$ and map how the energy landscape of all patterns evolve across the ABO$_3$ series.

Despite the useful insights on correlation between $t_G$, tilting patterns and relative importance of the orthorhombic phase within the \textit{Glazer Templates}, by construction it is still restricted to a set of idealised, perfectly periodic octahedral rotation patterns. However, its construction philosophy does allow one to incorporate more complex or mixed distortions, i.e. combination of IR/OR distortions around three different lattice directions and B-site displacements representable within larger cells.

%%%%%%%%%%%%%%%%%%%%%%%%%%%%%%%%%%%%%%%%%%%%%%%%%%%%%%%%%%%%%%%%%%%%%%%%%%%%%%%%%%%%%%%%%%%%%%%%%%%%%%%%%%%%%%%%%%%%%%%%%%%%%%%%%%%%%%%%%%%%%%%%%%%%%%%%%%%

\subsubsection*{Unconstrained Templates} 

\begin{figure*}[!htb]
\centering
\includegraphics{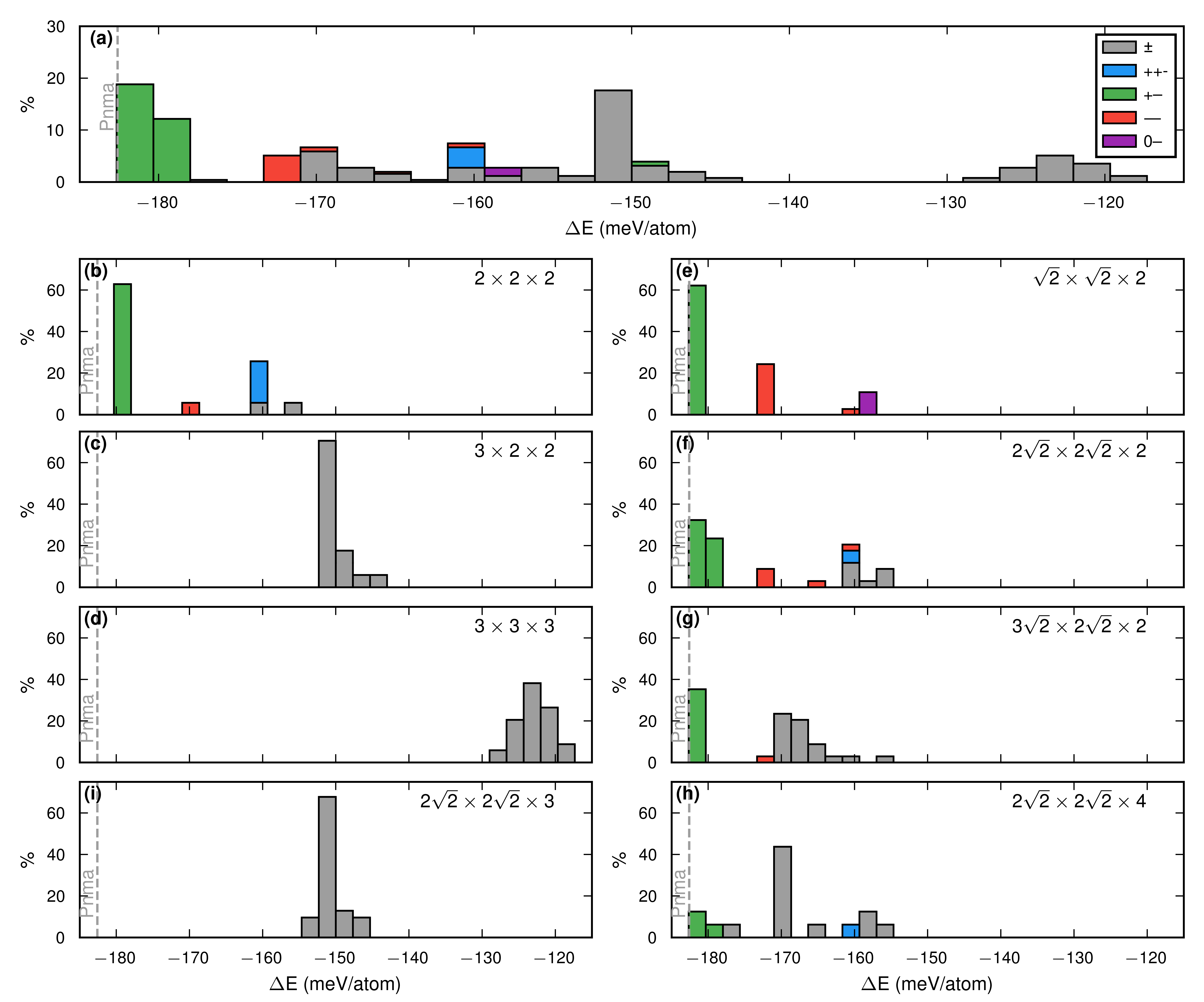}
\caption{\label{fig06:Distrb-Supercell}
Distribution of relative energies $\Delta E$ (meV/atom) for CaZrO$_3$ across all supercell templates with respect to cubic phase. Colours indicate different octahedral rotation patterns obtained from post-relaxation symmetry analysis: grey ($\pm$, mixed rotation), blue ($++-$), green ($+--$), red ($---$), and purple ($0--$). 
(a) All supercell configurations; (b–d) simple integer supercells $2 \times 2 \times 2$, $3 \times 2 \times 2$, and $3 \times 3 \times 3$; (e–g) rotated even supercells $\sqrt{2} \times \sqrt{2} \times 2$, $2\sqrt{2} \times 2\sqrt{2} \times 2$, and $3\sqrt{2} \times 2\sqrt{2} \times 2$; (h) rotated odd supercell $2\sqrt{2} \times 2\sqrt{2} \times 3$. The dashed vertical line marks the orthorhombic $Pnma$ ground-state energy.}
\end{figure*}

\textbf{Supercell Templates}\\
In this section we present the investigations on the exploration of all the \textit{Supercell Templates} (Sec.~\textit{Methods} and Sec.~S4(B1) of the SI), allowing both lattice vectors and internal coordinates to evolve freely. Structures in which the BO$_6$ units developed face or edge-sharing connectivity are classified as non-perovskite and removed from further analysis; these configurations are consistently higher in energy than the corner-sharing perovskite variants. For the remaining relaxed supercell structures, we analysed the octahedral rotation patterns by projecting the local rotations of each BO$_6$ octahedron onto the pseudo-cubic axes. To simplify notations of tilting patterns across different supercell templates, we drop the alphabet notations within the Glazer representation and only use the superscripts $0,+,-$. Following this nomenclature, we assign an effective Glazer-like symbol (e.g.\ $0--$, $+--$, $---$, $++-$, $+++$) whenever a consistent pattern extended across the supercell as shown in Fig.~\ref{fig06:Distrb-Supercell} (Fig.~S19 of the SI). In addition, we also encountered non-commensurate cases, i.e. the combination of tilt patterns does not tally with any of the Glazer notations. We grouped these non-commensurate cases as a separate mixed class, represented by the grey bars and marker ($\pm$) in Fig.~\ref{fig06:Distrb-Supercell}. A structural illustration of the $\pm$ pattern is provided in Fig.~S18 of the SI.

Here, we present the results of sampling \textit{Supercell Templates} for CaZrO$_3$, as shown in
Fig.~\ref{fig06:Distrb-Supercell}(a). Oganov-Valle structural fingerprint analysis of the 300 relaxed 
structures resulted in $\sim$250 unique supercell structures.  Among these supercell structures,  
the dominant tilt motifs are $\pm$ and $+--$ with 55.7 and 31.8\% contribution, respectively.
In contrast, the tilt patterns $---$, $++-$ and $0--$ only represent a minority with 7.0, 3.9 and 1.6\% contribution, respectively.

The $\pm$ class corresponds to configurations in which IR and OR coexist along at least one crystallographic axis, such that no single ideal Glazer notation can describe all BO$_6$ octahedra tilting pattern in the supercell. In contrast, we do not observe any configurations with $000$, $00+$, $00-$, $0++$, $0+-$, or the fully IR high-symmetry pattern $+++$. Their absence indicates that tilt systems involving one or more non-rotated axes, or purely IR along all three axes, are strongly disfavoured for CaZrO$_3$ under our sampling conditions. The only pattern containing a non-rotated axis that we identify is $0--$, which appears once (1.6\% of the dataset) in the $\sqrt{2} \times \sqrt{2} \times 2$ supercell at a relative energy of $\Delta E \approx -156$~meV/atom, as plotted in Fig.~\ref{fig06:Distrb-Supercell}(a) and (e) in purple. This rare appearance can be viewed as a metastable, relatively high-symmetry distortion in which rotations about one axis are effectively suppressed. It's rather low occurrence, despite the strong random perturbations applied to the initial structures, suggests that partially non-rotated or fully IR patterns are both rare in configuration space and energetically less favourable than the fully tilted $+--/---$ family.

The energy distributions in Fig.~\ref{fig06:Distrb-Supercell}(b)–\ref{fig06:Distrb-Supercell}(i)) reveal how the choice of supercell affects both the accessible tilt patterns and the quality of the low-energy sampling. All-even integer expansions of the primitive cubic cell, such as $2 \times 2 \times 2$ plotted in Fig.~\ref{fig06:Distrb-Supercell}(b), generate all frequently observed tilt motifs except $0--$, with relative energies spanning $-180 < \Delta E < -120$~meV/atom. Within this family, $+--$ and $\pm$ patterns dominate, with the $+--$ tilt pattern having $\Delta E \approx -179$~meV/atom, i.e.\ only $\sim 4$~meV/atom above the $Pnma$ ground state. This shows that $2 \times 2 \times 2$, and by extension larger supercells with even parity along all three directions, can realise a structure that is very close to the global minimum for CaZrO$_3$, and provide robust, general-purpose templates for structure searches.

In contrast, when an odd parity expansion is introduced along one or more lattice direction(s), as in $3 \times 2 \times 2$ (Fig.~\ref{fig06:Distrb-Supercell}(c)) and $3 \times 3 \times 3$ (Fig.~\ref{fig06:Distrb-Supercell}(d)), the situation changes qualitatively. Along any axis containing an odd number of BO$_6$ octahedra, an ideal OR ($-$) rotation pattern (BO$_6$ octahedra rotating in a clockwise/counter-clockwise/clockwise/$\dots$ sequence) cannot be propagated consistently through the corner-sharing network. As a result, the effective Glazer character along this direction is frustrated and manifests as either a net IR component ($+$), a non-rotated state ($0$), or a mixed character ($\pm$). For both $3 \times 2 \times 2$ and $3 \times 3 \times 3$ supercell, all relaxed structures fall into the $\pm$ category as shown in Fig.~\ref{fig06:Distrb-Supercell}(c) and Fig.~\ref{fig06:Distrb-Supercell}(d), respectively. The lower-energy peak positions of $3 \times 2 \times 2$ supercells as compared to $3 \times 3 \times 3$ can be addressed to additional accessible IR/OR tilting patterns in the crystallographic direction with even parity. Thus, integer supercells with odd parity along at least one axis not only constrains the apparent tilt signature along that axis to appear as “$+$, $0$ or $\pm$”; but, as the cell size increases, they also provide more internal degrees of freedom for the system to split into regions with different tilt patterns. This leads to a systematic increase in the fraction of $\pm$ with system size and to a corresponding shift of the energy distribution towards higher-lying, mixed-tilt structures, despite underlying chemical preference of CaZrO$_3$ towards a $+--$ tilt pattern.

We now focus on rotated-lattice supercells with even parity %along the $c$ direction
, $\sqrt{2} \times \sqrt{2} \times 2$, $2\sqrt{2} \times 2\sqrt{2} \times 2$, $3\sqrt{2} \times 2\sqrt{2} \times 2$, and $2\sqrt{2} \times 2\sqrt{2} \times 4$, as shown in Fig.~\ref{fig06:Distrb-Supercell}(e)–(h). These cells exhibit a relatively narrow energy distribution, typically $-182 < \Delta E < -155$~meV/atom, while still sampling the same dominant tilt motifs as the \textit{simple integer supercells}. Within this family, we identify the overall lowest-energy configuration of the entire CaZrO$_3$ dataset: a homogeneous $+--$ tilt pattern with $\Delta E \approx -183$~meV/atom, essentially degenerate with the orthorhombic ($Pnma$) reference. This shows that these rotated supercells with even parity are particularly well matched to the $Pnma$-type $a^+b^-b^-$ tilt network, simply because these supercells correspond to even integer expansion of the $\sqrt{2} \times \sqrt{2} \times 1$ $Pnma$ primitive cell. This interpretation also extends to the $2 \times 2 \times 2$ supercell (Fig.~\ref{fig06:Distrb-Supercell}(b)) which is related to $\sqrt{2} \times \sqrt{2} \times 1$ $Pnma$ primitive cell by a lattice basis transformation.

The statistics of tilt patterns within this rotated family also reveal how increasing system size modifies the expression of preferred tilt patterns. The fraction of structures with a $+--$ tilt pattern decreases systematically with supercell size: from 62.2\% in $\sqrt{2} \times \sqrt{2} \times 2$ to 55.9\%, 35.3\%, and 18.8\% in $2\sqrt{2} \times 2\sqrt{2} \times 2$, $3\sqrt{2} \times 2\sqrt{2} \times 2$, and $2\sqrt{2} \times 2\sqrt{2} \times 4$, respectively. Conversely, the proportion of $\pm$ mixed tilt configurations increases from 0.0\% to 23.5\%, 61.8\%, and 75.0\% over the same sequence. This trend reflects the additional internal degrees of freedom available in larger supercells. And, surprisingly the energy difference between the lowest energy $+--$ structure to that of the $\pm$ structure goes down gradually from $\sim$20 meV/atom in $2\sqrt{2} \times 2\sqrt{2} \times 2$, $\sim$10 meV/atom in $3\sqrt{2} \times 2\sqrt{2} \times 2$ to $\sim$5 meV/atom in $3\sqrt{2} \times 2\sqrt{2} \times 4$. This clearly illustrates that though the $+--$ framework acts as primary driving force in improving thermodynamic ranking, larger supercells enables mapping of multiple regions of mixed tilt patterns which can be thermodynamically competitive. As a result, these rotated even-parity supercells not only capture the true ground-state tilt pattern in CaZrO$_3$, but also naturally reveal a broader set of quasi-degenerate $\pm$ configurations close in energy to the ground state.

By contrast, rotated-lattice supercells with odd parity %along the $c$ direction
, $2\sqrt{2} \times 2\sqrt{2} \times 3$ as shown in Fig.~\ref{fig06:Distrb-Supercell}(i), are more restrictive in terms of clean Glazer patterns: all relaxed perovskite structures in this family (100\%) are classified as $\pm$, and their energies are shifted to higher values, roughly $-155 < \Delta E < -145$~meV/atom with respect to the cubic phase. This outcome can be understood from the perspective that in terms of parity, the $2\sqrt{2} \times 2\sqrt{2} \times 3$ cell is analogous to the $3 \times 2 \times 2$ integer supercell: both have an odd number of octahedra along one axis and even numbers along the other two axes. This odd parity along one direction frustrates an OR ($-$) pattern and forces the system towards $\pm$ or $+$/$0$ behaviour along the specific axis.

Overall, these observations indicate that both supercell shape, size and parity along each axis influences which tilt patterns can be realised at low energy: even-parity supercells favour coherent anti-phase networks close to the $Pnma$ ground phase, whereas large, odd-parity supercells tend to promote higher-energy mixed tilt configurations.

We further extend this set of supercell analyses to BaTiO$_3$ ($t_G = 1.071$) and SrTiO$_3$ ($t_G = 1.009$), omitting only the largest $2\sqrt{2} \times 2\sqrt{2} \times 4$ case. For BaTiO$_3$, the distribution of relative energies is very narrow, $0 < \Delta E \leq 15$~meV/atom (Fig.~S19(c) in SI), and almost all \textit{Supercell Templates} are able to describe the lowest-energy phase. Combined with the structural analysis as discussed later, our study confirms that, for BaTiO$_3$, B-site off-centering is the primary energy lowering mechanism, whereas changing the tilt pattern modifies the energy only weakly (within a few tens of meV/atom), in sharp contrast to the strongly tilt-driven CaZrO$_3$ case.

For SrTiO$_3$, we reused the relaxed perovskite CaZrO$_3$ geometries as initial templates by substituting the atomic species. The resulting energy spread is very small, on the order of $\sim 4$~meV/atom (Fig.~S19(b)in SI), again consistent with a very weak influence of octahedral tilts at $t_G \approx 1.00$. The dependence on \textit{Supercell Templates} follows the same qualitative trends as for CaZrO$_3$, i.e. even-parity cells more easily support coherent tilts than odd-parity ones. But, the absolute energy differences between tilt patterns are strongly reduced, indicating that SrTiO$_3$ thermodynamically prefers a cubic structure with only a small tendency to develop rotations of the BO$_6$ network.

Hence, the extensive analysis on \textit{Supercell Templates} leads to three main conclusions:
\begin{enumerate}
    \item For low-$t_G$ compositions such as CaZrO$_3$, random supercell sampling confirms that the $+--/---$ family of tilt patterns dominates the low-energy domain of potential energy surface and naturally reproduces $Pnma$ phases when the supercell is commensurate with the underlying tilt periodicity. In this case, supercell shape, size and parity directly control whether the system realises a uniform tilt pattern or breaks up into mixed-tilt ($\pm$) configurations.
    
    \item Supercells based on rotated $\sqrt{2} \times \sqrt{2} \times 1$ vectors with even parity along $c$ are particularly effective at accessing low-energy tilt phases, while simple integer expansions provide a more global but less focused sampling of both low- and high-energy structures.
    
    \item As $t_G$ $\geqslant$ 1.00 (SrTiO$_3$, BaTiO$_3$), the overall range of relative energies collapses, indicating a diminishing influence of collective octahedral tilts and an increasing propensity for cubic or only weakly distorted structures, with B-site off-centering becoming important only in specific compositions such as BaTiO$_3$. Depending on whether the objective is to explore the full energy landscape or to identify low-energy phases efficiently, different supercell templates can therefore be selected and combined with the symmetry constrained templates introduced earlier in this work, with the choice of template becoming most critical in the strongly tilt-driven, low-$t_G$ regime.
\end{enumerate}

\textbf{Distorted Templates}\\
Focusing on the results of the investigation of four different \textit{distorted templates}, we observe that the OR and IR tilting patterns via \textit{d-tetragonal OR} and \textit{d-tetragonal IR} templates, respectively have the strongest impact on improving thermodynamic ranking across all compositions and closely follows the orthorhombic phase ($+--$) as shown in Fig.~S20 of the SI. Due to the nature of the structure construction methodology, these templates also incorporate B-site displacement along with octahedral tilting patterns. The trend of $\Delta E$ vs $t_G$ for \textit{s-cubic} template closely follows other \textit{distorted templates}. Finally, the influence of B-site displacement represented by \textit{d-cubic} on lowering the thermodynamic ranking against the cubic phase is minimal.

In comparison to different \textit{Standard templates}, the trend lines of tetragonal-I ($P4mm$) and tetragonal-II ($P4/mbm$) lie in between the two trending lines of \textit{d-cubic} and \textit{s-cubic}. Details on the trends can be found in Fig.~S20(b) of the SI. The resulting relaxed lattice parameters, distortions and energetics of all the \textit{distorted templates} for all compositions can be found in Table.~S6, Sec.~S4(B2) of the SI. \\

%\subsubsection*{Influence of Distortions}
\textbf{Influence of Distortions}\\
Having probed the influence of the choice of different supercell shapes, size and parity on accessing low and high energy structures, we now explore how different structural distortions are expressed. To probe this, we introduce a set of inter-octahedral and intra-octahedral descriptors to map out distortions both between and within the BO$_6$ octahedron, respectively. As representative of inter-octahedral descriptors, we adopt the average sine of the octahedral tilting angle, $\langle \sin(\theta_{\text{tilt}}) \rangle$, which measures the amplitude of tilting of the corner-sharing BO$_6$ octahedra. As representative of intra-octahedral descriptors we adopt two physical quantities: (a) the  average of magnitude of the B-site off-centre displacement $\langle \Delta r_B \rangle$ (\AA) and (b) average of order parameter $\langle q \rangle$ characterising the amount of anisotropy within the octahedron following the spirit of local structural order parameters introduced in related contexts.~\cite{errington2001relationship}. Details on the mathematical definition and computation of these descriptors are provided in \textit{Sec.} \textit{Methods} and in Sec.~S3 of the SI.

\begin{figure*}[!thb]
\centering
\includegraphics{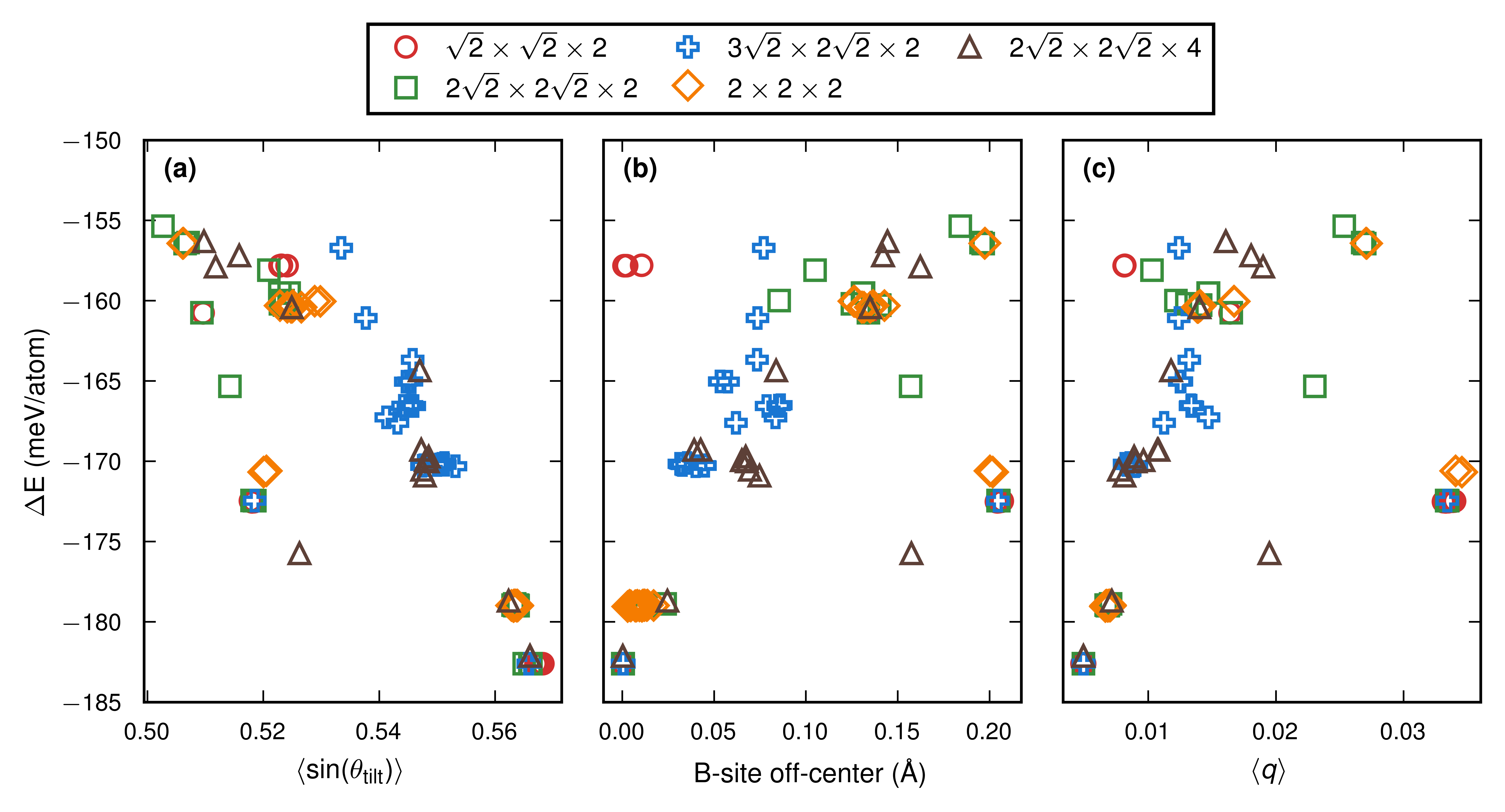}
\caption{\label{fig07:Feature-Supercell}
Correlation between relaxed energy $\Delta E$ (meV/atom) with respect to the cubic phase and structural descriptors for CaZrO$_3$ across different randomly distorted supercells. Each point corresponds to a relaxed structure, and the symbol indicates the supercell size. 
(a) Average tilting parameter $\langle \sin(\theta_{\text{tilt}}) \rangle$.
(b) Average B-site cation off-centre displacement $\Delta r_B$ (\AA).
(c) Average intra-octahedral order parameter $\langle q \rangle$.
}
\end{figure*}

The calculated descriptors $\langle \sin(\theta_{\text{tilt}}) \rangle$, $\langle \Delta r_B \rangle$, and $\langle q \rangle$ as functions of $\Delta E$ for different supercell templates of CaZrO$_3$ are shown in Fig.~\ref{fig07:Feature-Supercell}(a)–(c). For clarity, the supercell templates with odd parity are not shown here. Overall, the data fall into two main energy ranges. First, there is a cluster of low-energy structures around $\Delta E \approx -180$~meV/atom, predominantly generated from the $2\times2\times2$ and $\sqrt{2}\times\sqrt{2}\times2$ supercells. These structures exhibit relatively large values of $\langle \sin(\theta_{\text{tilt}}) \rangle$ (octahedral rotations with large amplitude) but almost vanishing intra-octahedral distortions: both $\langle  \Delta r_B \rangle $ and $\langle q \rangle$ are close to zero. This indicates that, for CaZrO$_3$, the ground-state stabilisation is mainly achieved through collective octahedral rotations, while B-site off-centring and anisotropy within octahedron are not required to reach the lowest energy configurations~\cite{dravid1989crystal,stoch2012crystal,feteira2008crystal}.

A second family of metastable structures appears around $\Delta E \approx -170$~meV/atom. In this regime, several points show simultaneous inter-octahedral and intra-octahedral distortions, with $\langle \Delta r_B \rangle$ reaching values of $\sim 0.15$–0.20~\AA{} and $\langle q \rangle \sim 0.02$, and are largely associated with the $3\sqrt{2}\times2\sqrt{2}\times2$ supercells (as demonstrated in Fig.~\ref{fig07:Feature-Supercell}). These structures are thermodynamically competitive within $\sim$10 meV/atom of the ground state; yet they rely on a cooperative interplay between octahedral rotations and B-site off-centring. Their presence illustrates that, even in a low-$t_G$ system where rotations dominate the absolute ground state, intra-octahedral distortions can still generate a rich set of thermodynamically competitive metastable phases.~\cite{bechtel2018octahedral,zhao2020polymorphous,yang2020assessment}

By contrast, \textit{Supercell Templates} with odd-parity display a relatively limited spread in their intra-octahedral descriptors compared to the inter-octahedral ones (Fig.~S21, Sec.~S4(C) in SI). For these supercells, $\langle \Delta r_B \rangle$ and $\langle q \rangle$ cluster around $\sim0.10$~\AA{} and $\sim0.02$, respectively, while $\langle \sin(\theta_{\text{tilt}}) \rangle$ remains confined to the range 0.45–0.52. Despite this non-negligible combination of distortions, their energies do not drop below $\Delta E \approx -150$~meV/atom. In other words, these large supercells explore a narrower region of the distortion space that does not intersect the global minimum identified with smaller supercells with all-even parity. Within the present sampling, this suggests that $3\times3\times3$ and $2\sqrt{2}\times2\sqrt{2}\times3$ based templates are incompatible with the lowest energy structures in CaZrO$_3$. This also confirms that efficient searches for low-energy configurations require supercells that are integer multiples of the $\sqrt{2}\times\sqrt{2}\times2$ cell,~\cite{xie_generation_2020,Lee_2016,emery2017high} and thereby able to capture all possible Glazer-type distorsions, including the $Pnma$ corresponding to the ground-state in the case of CaZrO$_3$.

For SrTiO$_3$ ($t_G = 1.009$), the descriptor–energy maps (Fig.~S22, Sec.~S4(C) in SI) show a very flat landscape: all relaxed supercell structures lie within about $\sim$5~meV/atom of the cubic reference and remain close to cubic. The lowest-energy phases,  obtained from $2\times2\times2$ and $\sqrt{2}\times\sqrt{2}\times2$ supercells, combine moderate octahedral rotations with finite Ti off-centring, indicating that octahedral rotations and B-site displacements contribute on a comparable footing.~\cite{he2022structural,evarestov2011phonon,loetzsch2010cubic,lan2019observing} For BaTiO$_3$ ($t_G = 1.071$), the plots in Fig.~S23, Sec.~S4(C) of the SI reveal that octahedral rotations are very small and vary only weakly, whereas $\langle \Delta r_B \rangle$ and $\langle q \rangle$ correlate strongly and almost linearly with $\Delta E$. In this high-$t_G$ limit, Ti off-centring within nearly rigid octahedra is therefore the dominant distortion mode, and the energy differences are mainly controlled by the magnitude of this local displacement.~\cite{merz1949electric,cohen1990lattice,cohen1992origin,acosta2017batio3,pal2020origin}

Extending this descriptor-based analysis to other ABO$_3$ compositions, we selected the three lowest-energy supercell templates of size $\sqrt{2} \times \sqrt{2} \times 2$ identified for CaZrO$_3$, SrTiO$_3$, and BaTiO$_3$ and used them as starting points to relax all 15 compositions. Their relative energies with respect to the ideal cubic phase, together with the corresponding values of $\langle \sin(\theta_{\text{tilt}}) \rangle$ and $\langle \Delta r_B \rangle$, are reported in Table~S7, Sec.~S4(C) of the SI. These template-driven tests leads to the key message that compositions with low $t_G$ tend to prefer CaZrO$_3$ like templates, characterised by large octahedral rotations and negligible B-site off-centring, whereas increasing $t_G$ gradually favours templates in which smaller rotations coexist with larger $\Delta r_B$. In other words, as $t_G$ increases, the dominant stabilisation mechanism shifts from collective octahedral rotations to more localized B-site displacements, while the overall energetic gain with respect to the cubic phase becomes smaller.

\begin{figure*}[!htb]
    \centering
    \includegraphics[scale=0.90]{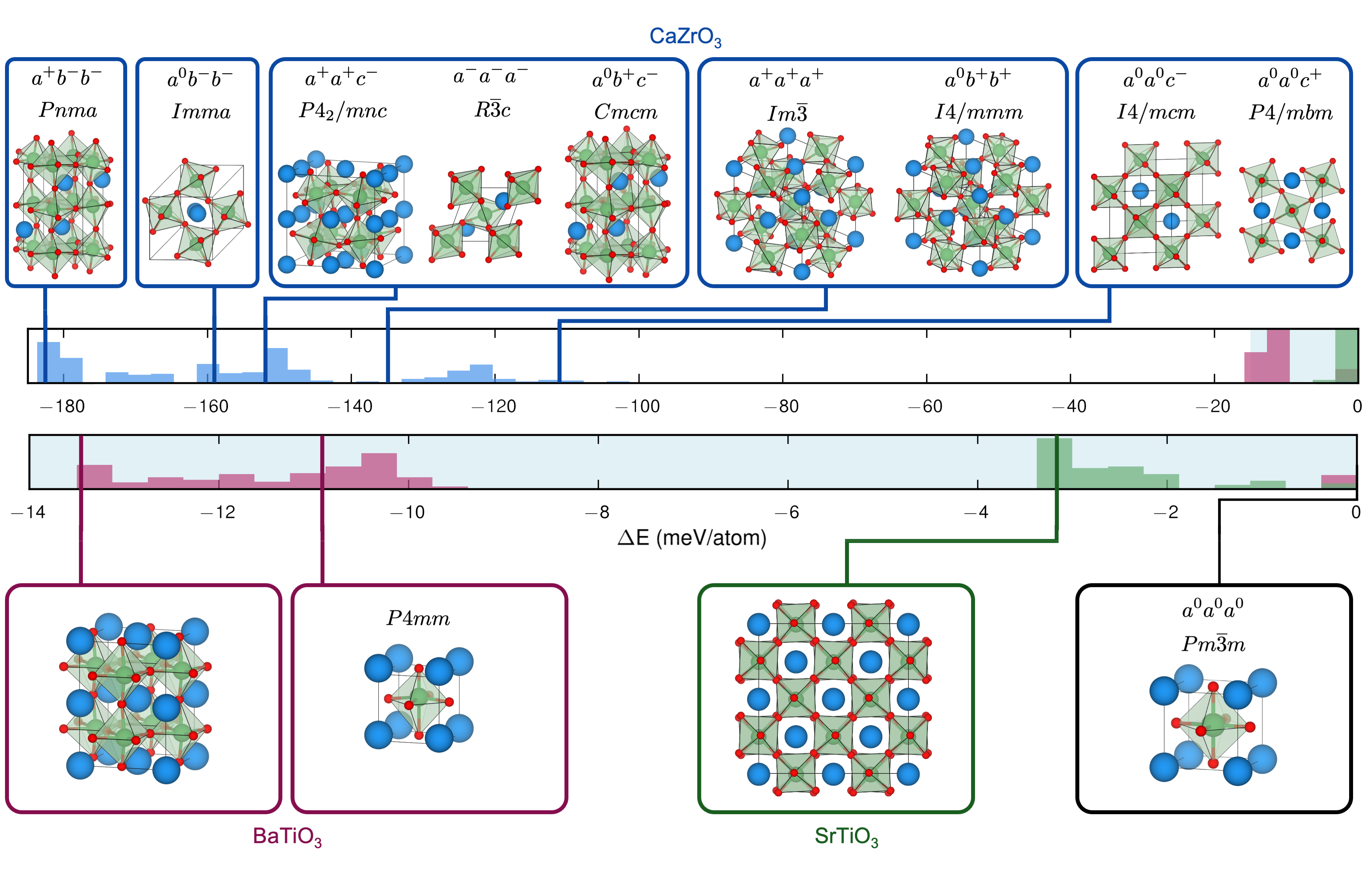}
    \caption{Thermodynamic energy landscape for CaZrO$_3$ (top), BaTiO$_3$ (bottom left), and SrTiO$_3$ (bottom right). The relative energies $\Delta E$ (meV/atom) of all optimised structures are shown with respect to the cubic $Pm\overline{3}m$ phase.}
    \label{fig08:Energy-Landscape}
\end{figure*}

Based on the extensive tests across different structural templates and compositions, we now construct a cumulative picture of the thermodynamic landscape for CaZrO$_3$ (top), SrTiO$_3$ (bottom right), and BaTiO$_3$ (bottom left), as shown in Fig.~\ref{fig08:Energy-Landscape}. The figure combines the energies of all 
\textit{Glazer}, \textit{symmetry constrained} and \textit{unconstrained} structural templates relative to the cubic $Pm\overline{3}m$ phase~\cite{glazer1972classification,howard1998group,simpson2025octahedral,xie_generation_2020,bechtel2018octahedral,zhao2020polymorphous}. The overall range of $\Delta E$ shrinks markedly with increasing $t_G$: CaZrO$_3$ spans nearly $\sim$200~meV/atom between the cubic reference and the ground state, SrTiO$_3$ displays a narrower spread ($\sim$4 meV/atom), and BaTiO$_3$ exhibits only modest stabilisation relative to the cubic phase ($\sim$15 meV/atom). Along this sequence, the characteristic nature of the lowest-energy distortions also alters: for small $t_G$ (CaZrO$_3$), strong inter-octahedral rotations dominate, whereas for large $t_G$ (BaTiO$_3$), intra-octahedral distortions and B-site off-centring become more prominent~\cite{dravid1989crystal,stoch2012crystal,evarestov2011phonon,he2022structural,acosta2017batio3}.

On delving deeper on the structural models of CaZrO$_3$, sorting the structures by energy reveals a sequence of increasingly complex octahedral rotation patterns as the energy decreases from the cubic phase towards the ground state. Simple one-axis tilt systems such as $a^0a^0c^-$ and $a^0a^0c^+$ already reduce $\Delta E$ to about –110~meV/atom. When tilt develops about both two axes and three axes, the energy is lowered further. Among different combinations, the most favourable arrangement corresponds to the Glazer $+--$ rotation pattern. This octahedral tilting pattern is realised in the orthorhombic ($Pnma$) phase, which emerges as the global minimum both among the \textit{symmetry-constrained Glazer templates} and among the \textit{supercell-derived structures}.

Similar behaviour has also been observed across most of the compositions considered here: where a perovskite ground state exists, $Pnma$ is typically the lowest-energy perovskite phase at 0~K within our set of \textit{Glazer} and \textit{standard templates}. Our \textit{ab-initio} findings are consistent with the widely reported experimental observation that many oxide perovskites adopt an orthorhombic ($Pnma$) structure at low temperature and transform to higher-symmetry (often cubic) phases only at higher temperature.

In summary, both inter-octahedral and intra-octahedral distortions act as key mechanisms for lowering the energy of perovskite oxides. For low-$t_G$ systems, large and correlated octahedral rotations are the primary driver of stabilisation and generate a wide landscape of low-energy phases, while intra-octahedral distortions only provide secondary refinements. For large-$t_G$ systems, by contrast, the tendency for octahedral rotations is significantly weaker and B-site off-centring within nearly rigid octahedra becomes the dominant route to stabilisation within the perovskite framework.

%%%%%%%%%%%%%%%%%%%%%%%%%%%%%%%%%%%%%%%%%%%%%%%%%%%%%%%%%%%%%%%%%%%%%%%%%%%%%%%%%%%%%%%%%%%%%%%%%%%%%%%%%%%%%%%%%%%%%%%%%%%%%%%%%%%%%%%%%%%%%%%%%%%%%%%%%%%
\subsection*{Influence of Temperature on Thermodynamic Ranking}
\label{subsec4:THERMODYNAMIC}
\begin{figure*}[!htb]
    \centering
    \includegraphics[scale=0.90]{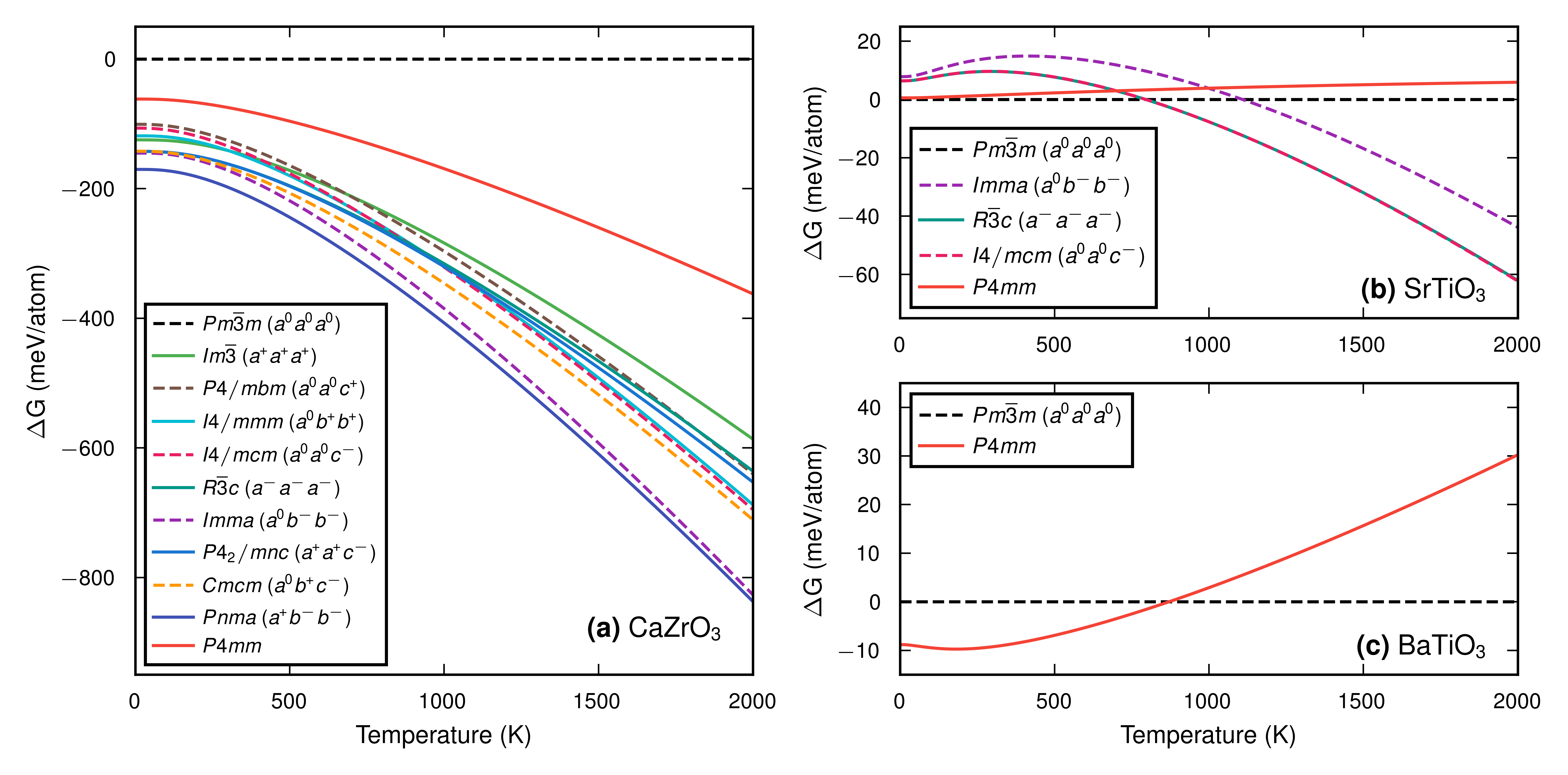}
    \caption{Temperature-dependent relative free energies $\Delta G(T)$ (meV/atom) for CaZrO$_3$ (top), SrTiO$_3$ (middle), and BaTiO$_3$ (bottom) with respect to the cubic phase (black dashed line). Coloured curves correspond to different symmetry phases in the legend.}
    \label{fig09:Free-Energy}
\end{figure*}

Extensive exploration of \textit{Symmetry-Constrained Templates} and \textit{Unconstrained Templates} across 15 representative perovskites compositions highlights the strong correlation between composition, nature of distortions and structural templates optimal for modeling these structural behaviour and how their suitability can be rapidly assessed via the $t_G$. Given that these DFT calculations are representative of thermodynamic ranking of different phases at 0~$K$, it becomes imperative to also assess their ranking as function of temperature via the computed free energy $\Delta G(T)$ taking into account contribution from phonons, limited to the quasi-harmonic approximation in this study. 
Details regarding the calculation of vibration spectra is provided in the methods section and in Sec.~S2 of the SI. As calculation of the vibration spectrum is computationally demanding, we limited our exploration to all distinct structures of CaZrO$_3$, SrTiO$_3$, and BaTiO$_3$ generated from the \textit{Glazer Templates}, supplemented with the tetragonal $P4mm$ phase representative of pure B-site distortion. After removing post relaxed similar structures for each compositions, we are left with 11, 5, and 2 candidate phases for CaZrO$_3$, SrTiO$_3$, and BaTiO$_3$, respectively. The calculated free energy for different structures within these three compositions are shown in Figure~\ref{fig09:Free-Energy}, as a function of temperature.

%%%%%%%%%%%%%%%%%%%%%%%%%%%%%%%%%%%%%%%%%%%%%%%%%%%%%%%%%%%%%%%%%%%%%%%%%%%%%%%%%%%%%%%%%%%%%%%%%%%%%%%%%%%%%%%%%%%%%%%%%%%%%%%%%%%%%%%%%%%%%%%%%%%%%%%%%%%

In the case of CaZrO$_3$, the orthorhombic ($Pnma$) phase emerges as the lowest-free-energy structure over the entire $T$=0–2000~K temperature range (left panel). Other distorted structures ($Imma$, $R\bar{3}c$, $Cmcm$, etc.) lie at higher $\Delta G(T)$ closely following the $Pnma$ phase. All these phases are separated by several hundred meV/atom across the full temperature window with respect to the cubic Pm$\bar{3}$m phase. Within the present set of tilt patterns, our calculations therefore do not predict any temperature-driven phase transition for CaZrO$_3$ up to 2000~K: $Pnma$ remains the thermodynamic ground state, and the free-energy landscape is dominated by low-symmetry octahedral rotations.~\cite{dravid1989crystal,stoch2012crystal,de1996high,feteira2008crystal}

Experimentally, SrTiO$_3$ has been reported to undergo a phase transition from the anti-ferrodistortive tetragonal phase ($I4/mcm$) to cubic phase  $Pm\overline{3}m$ at transition temperature of $T$=105~K \cite{okazaki1985high}. And, this picture is consistent with the general perception that perovskite oxides adapt low symmetry distorted phases at low temperatures and sustain the high symmetry cubic phases only at high temperature. However, this generic and experimental observation is inconsistent with the computed free energy profile of different phases, where the cubic $Pm\overline{3}m$ phase emerges as most stable with $\sim$10 meV/atom at $T$=0$K$  and remains lowest in free energy up to around $T \approx 800$~K as shown in Fig.\ref{fig09:Free-Energy}(b). Above this temperature, both the rhombohedral $R\overline{3}c$ phase and the tetragonal $I4/mcm$ phase become slightly more stable than $Pm\overline{3}m$ and remain essentially degenerate with each other over the explored temperature range. The $Imma$ phase is close in free energy and also becomes slightly more favourable with temperature, while the $P4mm$ and $Pm\overline{3}m$ phases stay nearly degenerate over the full range.

Based on our calculations within harmonic approximations, the free-energy surface of SrTiO$_3$ is very flat: apart from the weak crossover between $Pm\overline{3}m$ and the nearly degenerate $R\overline{3}c$ and $I4/mcm$ pair, no additional clear phase transitions appear within the limited set of templates considered here. Instead, several low-symmetry tilt systems remain clustered within a narrow free-energy window of the cubic reference, consistent with the well-known incipient and competing rotational instabilities in SrTiO$_3$.~\cite{evarestov2011phonon,he2022structural,loetzsch2010cubic,lan2019observing}. In addition, SrTiO$_3$ has been known to be a prototypical case with strong anharmonicity and necessitates to go beyond the harmonic approximation to account for correct free energy profile between different competing phase\cite{he2020anharmonic}, which is beyond the scope of this work.

For BaTiO$_3$, we restrict our analysis to the cubic $Pm\overline{3}m$ and tetragonal P4mm phases (bottom right panel). 
Among the two phases the P4mm phase has a lower free energy than $Pm\overline{3}m$ by $\sim$10 meV/atom at low temperature. As temperature increases, the difference between the two phases decreases. At intermediate temperature of $\sim$880~K, the free energy curve of $P4mm$ goes above that of the $Pm\overline{3}m$ phase, which indicates that our harmonic free-energy calculations predict a tetragonal-to-cubic transition for BaTiO$_3$, with P4mm stable at low $T$ and $Pm\overline{3}m$ stabilised by vibrational entropy at high $T$
~\cite{cohen1990lattice,merz1949electric}. Though the trend of phase transition is consistent with experimental observation,
the computed temperature of $\sim$880~K is much higher than the the experimentally reported transition temperature $T$=398~K\cite{acosta2017batio3}. Similar to SrTiO$_3$, it is also necessary to account for anharmonic effects in BaTiO$_3$ for correct quantitative description of free energy profiles and phase transitions~\cite{eklund2023pyroelectric}.

Overall, these results highlight how the same set of symmetry-breaking distortions generates very different temperature dependent free-energy profiles across the three compositions: a strongly stabilised orthorhombic ground state with no phase change in CaZrO$_3$, a shallow multi-minima landscape in SrTiO$_3$, and a well-defined tetragonal-to-cubic crossover in BaTiO$_3$. Within the limitations of the harmonic approximation and the restricted set of structural templates considered here, the computed $\Delta G(T)$ provides a coherent picture of how octahedral rotations and B-site off-centre displacements compete as a function of temperature and $t_G$, while absolute transition temperatures and narrow stability windows should be regarded as approximate.

%##############################################################################

\subsection*{Physical Properties}
\label{subsec5:PHYPROPERTIES}
Beyond the influence of distortions such as octahedral rotations and B-site displacements on thermodynamic ranking of different perovskite phases, they are also known to have strong implications on various physical properties relevant for real-world applications. Hence, in addition to our systematic thermodynamic exploration of different structural templates, we also investigate the electronic and mechanical properties of these systems in terms of the HOMO–LUMO gap $E_g$ (eV) and bulk modulus $B_0$ (GPa), respectively. The detailed methodology used to compute these quantities are provided in Sec.~S5(B–C) of the SI.

To evaluate how these physical properties vary with structure, we focused on 24 templates constructed by combining the \textit{Standard Templates} (5), \textit{Glazer Templates} (15) from the \textit{Symmetry Constrained Templates} and \textit{Distorted Templates}(4) from the \textit{Unconstrained templates} and computed $E_g$ and $B_0$ for all 15 compositions. The contour plots of resulting values of $E_g$ and $B_0$ across different structural templates are provided in Fig.~S24, Sec.~5(A) of the SI. And, corresponding values obtained for cubic phase and lowest energy phase are reported in Table~S8 of the SI, where lowest energy phase is either the orthorhombic phase or other phase as determined by the protocol as described in Sec.\ref{sec:PROTOCOL}.

\begin{figure*}[!htb]
\centering
\includegraphics[scale=0.90]{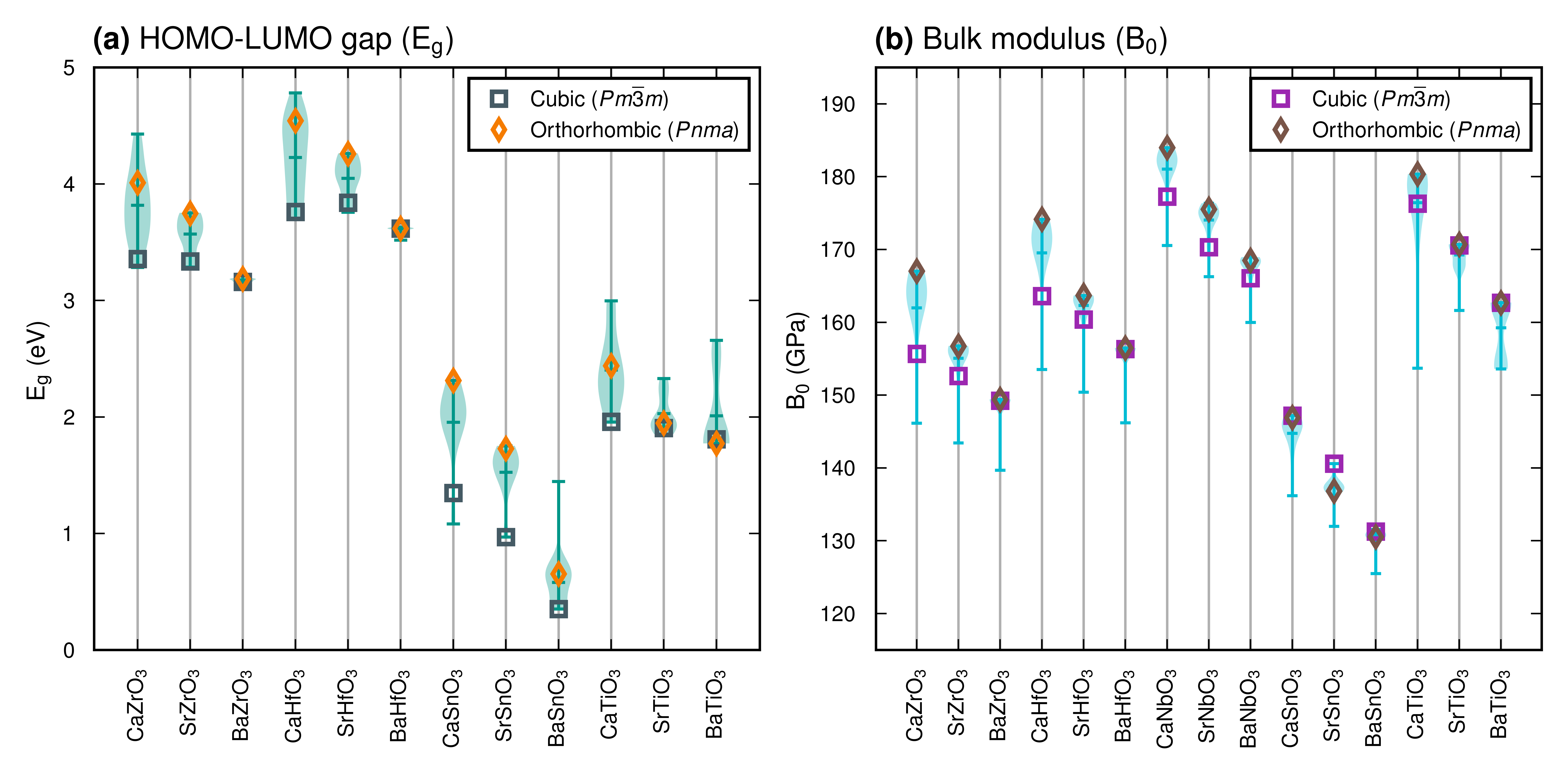}
\caption{\label{fig10:Properties-Violin} 
Violin plots of the calculated (a) HOMO–LUMO gaps obtained with PBE, $E_g$ (eV), and (b) bulk modulus $B_0$ (GPa), for all 15 compositions (x-axis) across the structural templates considered in this work. We removed the entries for ANbO$_3$ in (a) as they displayed metallic behaviour. For each composition, the top and bottom ends of the violin and the central line indicate the minimum, maximum, and average values, respectively. Cubic and orthorhombic phases are highlighted by grey open squares and orange open diamonds, respectively.}
\end{figure*}

The spread of $E_g$ and $B_0$ values across different templates is depicted by the violin plots in Fig.~\ref{fig10:Properties-Violin} (details on the calculation and values of $E_g$ for different composition and phases are provided in Table.~(S9,S10,S11) in Sec.~S5(B) of the SI). All compositions display consistent behaviour, particularly across the change of A-site cation with the exception of Sn and Ti based perovskites. The spread in the $E_g$ for each composition can be associated with both the composition and octahedral distortions. Among the \textit{Glazer Templates}, $Pm\bar{3}m$(open square) and $Pnma$(open diamond) phases possess the minimum and maximum $E_g$ values, respectively as shown in Fig.~\ref{fig10:Properties-Violin}(a). Most of the additional points beyond this range arise from the \textit{distorted templates} and are drastically different from the former case. While \textit{Glazer Templates} are representative of different octahedral rotations, \textit{distorted templates} constitute different combinations of IR, OR and B-site displacements. This implicit difference in the nature of displacement most likely lead to different trend in $E_g$ values.
Their combined influence via different templates leads to an overall spread in the $E_g$ from $\sim$1 eV for A=[Ca] to $\leqslant$0.5 eV for A=[Sr, Ba]; holding true for both the individual templates. It is also particularly interesting to note that the difference in $E_g$ between $Pm\bar{3}m$ and $Pnma$ decreases and closes while moving across Ca, Sr, Ba for a given B-site element. This can be associated with the decreasing role of distortion in thermodynamic stabilization in composition with high $t_G$ values such that $Pm\bar{3}m$ and $Pnma$ (and in fact all other \textit{Symmetry Constrained templates}) are structurally increasing similar to each other as $t_G$ increases and tilt angles get closer to 180$^0$. 

Unlike other compositions, Sn and Ti based perovskites present a different scenario on the values of $E_g$. These compositions depict a qualitatively similar behaviour for the \textit{Glazer Templates} with a decrease in the difference of $E_g$ values between $Pm\bar{3}m$ and $Pnma$ phases with increase in size of A-site cation. However, their $E_g$ distribution calculated from the \textit{distorted templates} extends far beyond the limits of $Pm\bar{3}m$-$Pnma$ $E_g$ range. In these cases, the B-site element displays a relatively stronger influence of unusual set of distortions on $E_g$; particularly the B-site displacement in the $P4mm$ phase. At the atomic level, such behaviour can be associated with the relative shift of the band edges and changing contribution of the Ti-$d$ and Sn-$p$ orbitals on these band edges along with different distortions~\cite{cohen1990lattice,linnik2021raman,maul2015silico,glerup2005high,chahib2023investigation}.

On a broader perspective, our representative ABO$_3$ systems display increasing $E_g$ with lowering symmetry, as reported for different extensively studied halide perovskites~\cite{matheu2022halogen}. Analysis of the trends reveal a relatively complex combined influence of both composition chemistry and octahedral distortions. While A-site cation is representative of a key to tune \textit{Glazer}-like octahedral rotations, the implicit nature of the B-site elements is representative of imparting relatively stronger or weaker influence on $E_g$. Hence, this can be interpreted as a combined influence of both ionic sizes of different chemical species impacting final structure and interplay of their chemical nature via orbital bonding.

In a similar fashion, the calculated $B_0$ bulk moduli show similar trends across varying A-site elements for a given B-site element, with the exception of Ti and Sn. Among  \textit{Glazer template} based models, the $Pm\bar{3}m$ phase has the minimum value and $Pnma$ phase the maximum. The gap between them closes with increasing size of A-site cation, as shown in Fig.~\ref{fig10:Properties-Violin}(b); consistent with reducing influence of octahedral tilting in large $t_G$ systems. The large spread in the $B_0$ can be associated with the \textit{distorted templates}. Unlike, the influence of the symmetry lowering mechanisms of \textit{distorted templates} in increasing $E_g$, these lead to decrease in $B_0$ values for all cases. While certain low symmetry IR/OR distortions lead to increasing $B_0$, others such as B-site displacement reduces $B_0$ with respect to the cubic phase. And, across different compositions and structural models, the variation in $B_0$ is only within $\sim$10-20\%. Calculation values of $B_0$ for different composition and phases are provided in Table.~(S12,S13,S14) in Sec.~S5(C) of the SI.

%%%%%%%%%%%%%%%%%%%%%%%%%%%%%%%%%%%%%%%%%%%%%%%%%%%%%%%%%%%%%%%%%%%%%%%%%%%%%%%%%%%%%%%%%%%%%%%%%%%%%%%%%%%%%%%%%%%%%%%%%%%%%%%%%%%%%%%%%%%%%%%%%%%%%%%%%%%
\subsection*{\label{sec:PROTOCOL}Optimisation Protocol and Validation}
Based on the computational experiments performed on different structural templates, %\ref{sec:STRUCTEMPLATES}
we find that structures containing a $+--$ octahedral rotation pattern, i.e. orthorhombic $Pnma$ phase on expanded supercells based on $\sqrt{2} \times \sqrt{2} \times 2$ unit cell systematically yield the lowest energies, self adjusting the extent of octahedral distortions for the given composition. This unit cell also manages to sample other thermodynamically competitive metastable phases with other tilt patterns. These findings addresses one of the main aims of this work, i.e. to quantify how different distortions in supercells of different shape, size and expansion parity affect sampling of thermodynamically relevant phases with different symmetries across 15 perovskite oxide compositions. Utilizing these emerging trends, we extended our effort towards design of a simple adaptable practical workflow which can easily generate suitable low symmetry structural models.

Following systematic tests across eight \textit{Supercell Templates} for CaZrO$_3$, SrTiO$_3$, and BaTiO$_3$, we find that a minimum supercell of effective size $\sqrt{2} \times \sqrt{2} \times 2$ is sufficient to sample both putative global minima and thermodynamically relevant low energy structures (details provided in Table.~S5, Sec.~4(B1) of the SI).  As an initial test on the robustness of the $\sqrt{2} \times \sqrt{2} \times 2$ unit cell, we first adapt the lowest-energy structural models obtained for CaZrO$_3$, SrTiO$_3$, and BaTiO$_3$ as prototypes on all the 15 compositions and evaluate their relative energetic ranking post relaxation.

\begin{figure*}[!htb]
\centering
\includegraphics[scale=0.16]{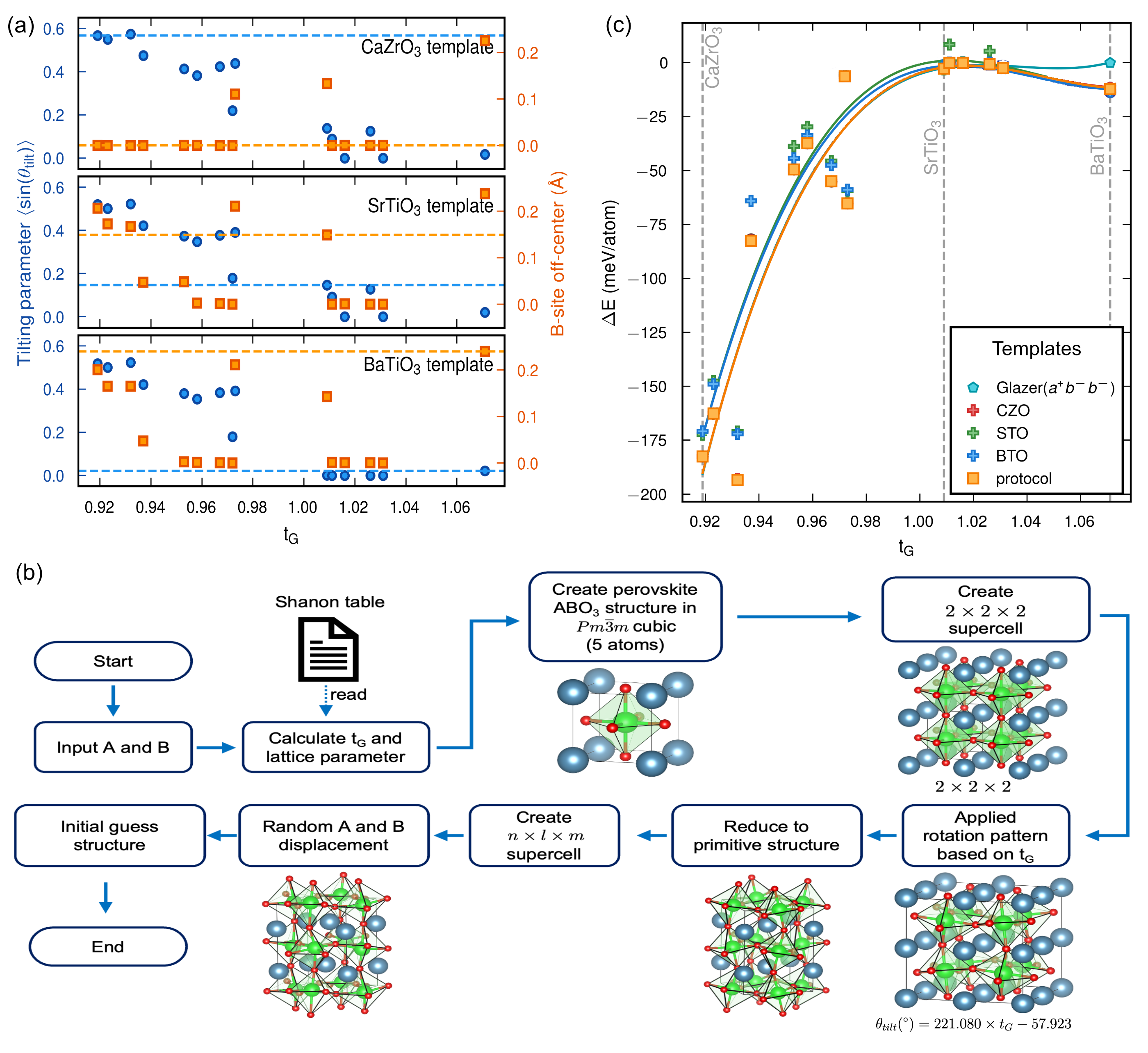}
\caption{\label{fig11:Protocol-Workflow}
(a) Tilt parameter $\langle \sin(\theta_{\mathrm{tilt}}) \rangle$ and B-site off-centering $\langle \Delta r_B \rangle$
plotted as a function of the tolerance factor $t_G$ for the 15 compositions. Dashed lines indicate the corresponding values before structural relaxation.
(b) Workflow for generating initial low symmetry distorted perovskite structures for \textit{ab initio} calculations.
(c) Relative energies $\Delta E$ (meV/atom) of all compositions with respect to the cubic phase, plotted as a function of $t_G$. The lowest-energy phases obtained from distorted-cubic (d-cubic), super-cubic (s-cubic), Glazer ($a^+b^-b^-$), supercell CaZrO$_3$ (CZO), SrTiO$_3$ (STO), BaTiO$_3$ (BTO), and protocol-based approaches are shown with different markers. In all cases, the protocol-generated structures provide the lowest-energy phase.}
\end{figure*}

Variations in key structural descriptors across the CaZrO$_3$, SrTiO$_3$, and BaTiO$_3$ derived templates for the full set of 15 compositions post relaxation, are shown in Fig.~\ref{fig11:Protocol-Workflow}(a) as a function of $t_G$ of the corresponding ABO$_3$ composition. Dashed lines indicate the corresponding structural parameters of the template prior to relaxation for comparison. Their relative energies with respect to the cubic phase, average tilt parameter $\langle \sin(\theta_{\mathrm{tilt}}) \rangle$ and the magnitude of the B-site off-centering $\langle \Delta r_B \rangle$ are listed in Table~S7, Sec.~S4(C) of the SI. A clear trend emerges for the inter-octahedral descriptor: $\langle \sin(\theta_{\text{tilt}}) \rangle$ shows a quasi-linear, monotonically decreasing correlation with $t_G$ across all three template families. By contrast, $\langle \Delta r_B \rangle$ does not display any correlation with $t_G$, which suggests that prescribing a fixed B-site off-centre displacement based on the composition alone is not reliable. Hence, in our protocol we use $t_G$ only to set the initial rotation amplitude, while B-site displacements are introduced as small random perturbations for generating initial structural models.

The protocol for identifying the low symmetry thermodynamically relevant (meta)stable perovskite structures is summarised schematically in Fig.~\ref{fig11:Protocol-Workflow}(b), and the individual steps are described below. The data in Fig.~\ref{fig11:Protocol-Workflow}(a) provide the initial points used to parameterise a simple composition-dependent model $t_G \rightarrow \theta_{\mathrm{tilt}}$ (Sec.~S6 in SI), which predicts the starting octahedral rotation angle $\theta_{tilt}$ for a given composition.

The workflow is general in the sense that any desired octahedral rotation pattern can be used as the starting point (for example $+--$, $0--$, or $0+-$); here we adopt $+--$ tilting pattern to demonstrate functionality of the workflow.

\begin{itemize}
\item \textbf{Step 1:} For a given ABO$_3$ composition, compute the $t_G$ from Shannon ionic radii and use it to estimate the lattice constant of the ideal cubic phase. Construct the cubic unit cell with space group $Pm\overline{3}m$.

\item \textbf{Step 2:} Build a simple integer $2 \times 2 \times 2$ expansion supercell from this cubic unit cell. At this stage, the structure is still undistorted and retains the symmetry of the cubic reference.

\item \textbf{Step 3:} Choose a target octahedral rotation pattern (here $+--$). Apply this pattern along the three different crystallographic axes and determine the smallest primitive cell that realises it. For the $+--$ pattern, this primitive cell has effective dimensions $\sqrt{2} \times \sqrt{2} \times 2$ with respect to the cubic lattice. The initial tilt angle $\theta_{\mathrm{tilt}}$ is set by a linear model as a function of $t_G$ (eq.~(S12) in the SI) and mentioned in Fig.~\ref{fig11:Protocol-Workflow}(b). This primitive distorted cell serves as the basic building block for subsequent supercell expansions.

\item \textbf{Step 4:} Depending on the desired system size and the distortions to be sampled, construct an integer supercell expansion $n \times l \times m$ of the primitive distorted cell. In this way, the protocol is flexible with respect to both the rotation pattern and the supercell shape and size.

\item \textbf{Step 5:} Within the chosen supercell, randomly displace the A and B site cations with a user defined amplitude to introduce additional local distortions. In the present work, we use a displacement amplitude of $|\Delta \vec{r}| = 0.2$~\AA. These displacements are not tuned by composition; instead, they provide a generic way to seed possible B-site off-centering that is then refined during relaxation.

\item \textbf{Step 6:} Using the resulting distorted structure model as the initial configuration, a full structural relaxation with an appropriate \textit{ab initio} method is carried out. This yields a relaxed structure and its relative energy.
\end{itemize}

To validate the workflow, we applied this protocol independently to all 15 compositions and benchmarked the resulting energies against those obtained from the full suite of \textit{Symmetry-Constrained Templates} and recent lowest energy $2 \times 2 \times 2$ \textit{Supercell Templates} obtained from CaZrO$_3$, SrTiO$_3$ and BaTiO$_3$. The corresponding comparison is reported in Fig.~\ref{fig11:Protocol-Workflow}(c).

With the lowest energy $2 \times 2 \times 2$ \textit{Supercell Templates}, the relaxed structures for all compositions (filled cross symbols) reach energies that are nearly as low as those of the most favourable Glazer $a^+b^-b^-$ template. 
But, by contrast the structural models generated by our proposed protocol subsequently relaxes (orange squares) as most stable candidates in all the cases.

These results explicitly demonstrate that with appropriate supercell templates, incorporating physically motivated octahedral rotation patterns in combination with a simple composition–tilt model and generic local displacements, can reliably identify low-energy phases while keeping the number of required relaxations modest. The protocol is flexible and adaptable with respect to both the chosen rotation pattern and the supercell shape and size; thus, enabling easy adaptation for exploration of different low and high symmetry tilt systems. As such, this workflow provides an robust, efficient and general starting point for high-throughput exploration of complex perovskite energy landscapes; particularly in the low $t_G$ regime where the thermodynamics is strongly rotation driven.

%%%%%%%%%%%%%%%%%%%%%%%%%%%%%%%%%%%%%%%%%%%%%%%%%%%%%%%%%%%%%%%%%%%%%%%%%%%%%%%%%%%%%%%%%%%%%%%%%%%%%%%%%%%%%%%%%%%%%%%%%%%%%%%%%%%%%%%%%%%%%%%%%%%%%%%%%%%
\section*{\label{sec:CONCLUSION}Conclusion}
In conclusion, our systematic exploration of different structural distortions across 15 representative ABO$_3$ perovskite oxides and multiple classes of templates reveal a coherent picture linking crystal structure and distortion, composition, and thermodynamic stability. 
%Point (i)
The extensive \textit{ab initio} studies establish a clear quantitative correlation between the Goldschmidt tolerance factor ($t_G$) and the propensity of a perovskite composition to stabilise in a low-symmetry phase relative to the cubic $Pm\overline{3}m$ reference via relative energy $\Delta E$: low $t_G$ systems exhibit large stabilization energies upon distortion, whereas high-$t_G$ systems remain closer in energy to the cubic phase. 

%Point (ii)
Systematic analysis of structural descriptors indicate the both inter-octahedral (tilts) and intra-octahedral (local distortions, B-site off-centring) modes emerge as key energy lowering mechanisms for low symmetry perovskite phases. While octahedral tilting is the key driver of stabilization for low $t_G$ compositions, B-site displacements combined with other intra-octahedral distortions becomes relevant for compositions with high $t_G$.

%Point (iii)
Within out set of representative perovskite compositions, the structural template with $+--$ tilting pattern appears as the most thermodynamically favourable low symmetry phase and hence, can serve as an excellent proxy model for identification of putative ground state configurations. Appropriate composition dependent B-site displacements can be further adapted on these structural templates for refined searches of low symmetry phases.  

%Point (iv)
We categorically demonstrate that irrespective of the number of randomized initial configurations, the choice of supercell shape , size and parity, commensurate with the nature of distortion in low symmetry phases plays the most crucial role in sampling low energy phases of perovskite oxides. On this basis, we have developed and validated a simple and adaptive optimisation protocol that is capable of generating physically relevant low symmetry structural models with appropriate nature and amount of distortions across different ABO$_3$ compositions.

%Point (v)
Finally, the implications of different structural distortions go beyond the thermodynamic stabilization of low symmetry phases. With increase in temperature, the relevance of these distortions diminishes as shown by the computed free energy as function of temperature. Compositions with $t_G$ $<$ 1.00 are unlikely to stabilize in high symmetry phases even at high $T$. In retrospect, compositions with $t_G$ $\approx$ 1.00 stabilises with high symmetry phases, particularly cubic.

However, compositions with $t_G$ $>$ 1.00 prefer to stabilise with phases constituting Jahn-Teller like distortions ($\Delta r_B$ $>$ 0) over the cubic phase at high $T$ or adapt non-perovskite phases. In parallel to the emergence of different phases with changing temperature, these distortions also induce significant change in the physical properties such as the HOMO-LUMO gap and the bulk modulus. Thus, distortions that minimise the total (or free) energy simultaneously also govern electronic structure and mechanical stiffness in a $t_G$-dependent manner. Understanding of this evolving structure-property relationship with temperature is critical for optimal design of materials for different applications.

Therefore, these results provide a unified computational framework for development of workflows enabling rapid exploration of both high and low symmetry perovskite structures with different symmetry breaking mechanisms. Detailed analysis of the inter-octahedral and intra-octahedral descriptors of the structures provide an easier conceptual access on understanding synergistic influence of composition-distortion on their thermodynamic and physical properties. The availability of such a physically grounded, adaptive and robust protocol for generating low symmetry structures with different distortions opens up opportunities for efficient large-scale high-throughput screening of new thermodynamically relevant perovskite compositions. Hence, these activities have potential to enrich existing materials databases with energetically relevant low symmetry perovskite models and thereby alleviate the current bias towards ideal cubic phase in open-source repositories.
 
%%%%%%%%%%%%%%%%%%%%%%%%%%%%%%%%%%%%%%%%%%%%%%%%%%%%%%%%%%%%%%%%%%%%%%%%%%%%%%%%%%%%%%%%%%%%%%%%%%%%%%%%%%%%%%%%%%%%%%%%%%%%%%%%%%%%%%%%%%%%%%%%%%%%%%%%%%%

\section*{Methods}\label{sec:COMPDETAILS}

\subsection*{DFT calculations}
Density functional theory (DFT) calculations across different compositions were performed using the Quantum ESPRESSO (QE) package \cite{giannozzi2009quantum,giannozzi2017advanced} with the generalized gradient approximation (GGA) of the 
Perdew
–Burke–Ernzerhof (PBE) exchange–correlation functional \cite{perdew1996generalized}. The elements were represented using Optimized Norm-Conserving Vanderbilt (ONCV) pseudopotentials \cite{hamann2013optimized,van2018pseudodojo}. Details pertaining to the setup of the kinetic energy cutoff, reciprocal-space sampling, convergence parameters, and other aspects are provided in Sec.~S2 of the SI.

\subsection*{Free Energy: Vibrational Properties}
In order to compute free energy of different phases, the phonon spectrum is computed within density functional perturbation theory (DFPT) as implemented in \textsc{QE}. The resulting dynamical matrices are post-processed with \textsc{Phonopy}~\cite{phonopy-phono3py-JPCM,phonopy-phono3py-JPSJ} to obtain the vibrational contribution to the Helmholtz free energy $F_{\mathrm{vib}}(T)$ within the harmonic approximation. The total Helmholtz free energy is then expressed as $G(T) = E_{\mathrm{DFT}} + F_{\mathrm{vib}}(T)$ in (meV/atom) with respect to the cubic $Pm\overline{3}m$ phase. 

Additional details on the details and set up of the calculation is provided in Sec.~S2 of the SI.

\subsection*{Structural descriptors}
As discussed in the introduction, perovskites primarily display two kinds of structural distortion: octahedral tilting between two adjacent octahedron and displacement of the B-site atom within an octahedron. To characterize these changes, we used the following two classes of descriptors: (a) \textit{inter-octahedral distortions} via the tilting parameter $\langle \sin (\theta_{tilt}) \rangle$, obtained by averaging the sine function across all the tilting angles $\theta_{tilt,i}$ as shown in eq.~(\ref{eq1:PARAMETER}), and (b) \textit{intra-octahedral distortions} via the order parameter $\langle q \rangle$, as shown in eq.~(\ref{eq2:order}), based on the $\angle$ O$_i$BO$_j$ ($\theta_{ikj}$)  angles within a single octahedron. We also introduced another parameter, $\langle \Delta r_B \rangle$ as shown in eq.~(\ref{eq3:delRB}), indicating the displacement of the B-site atom from the centroid within the octahedron $\vec{r}_{centroid}$ as shown in eq.~(\ref{eq4:OctCenter}). Instead of using the coordinates of the B-site atom of two adjacent octahedron to determine $\theta_{tilt}$ and tilting parameter, we used the $\vec{r}_{centroid}$ coordinates.

Within this framework, an ideal cubic phase with no tilting and no B-site displacement is represented by $\langle \sin (\theta_{tilt}) \rangle = 0$, $\langle q \rangle = 0$, and $\langle \Delta r_B \rangle = 0$. 
Details on the descriptors and their evaluation are provided in Sec.~S3 of the SI.

\begin{equation}\label{eq1:PARAMETER}
\langle sin (\theta_{tilt}) \rangle = \frac{1}{N}\sum_{i=0}^N sin (\theta_{tilt,i}) 
\end{equation}

\begin{equation} \label{eq2:order} 
\langle q \rangle = \frac{1}{N_B} \sum_{k=1}^{N_B} \left| 1 - \frac{3}{8} \sum_{i>j} \left( \cos(\theta_{ikj}) + \frac{1}{3} \right)^2 \right| 
\end{equation}

\begin{equation} \label{eq3:delRB} 
\langle \Delta r_{B} \rangle = \frac{1}{N_B} \sum_{k=1}^{N_B} \left| \vec{r}_{B,k} - \vec{r}_{centroid,k} \right| \\  
\end{equation}

\begin{equation} \label{eq4:OctCenter} 
 \vec{r}_{centroid,k} = \frac{\sum_{i=1}^6 \vec{r}_{O_{i,k}}}{6} 
\end{equation}

%%%%%%%%%%%%%%%%%%%%%%%%%%%%%%%%%%%%%%%%%%%%%%%%%%%%%%%%%%%%%%%%%%%%%%%%%%%%%%%%%%%%%%%%%%%%%%%%%%%%%%%%%%%%%%%%%%%%%%%%%%%%%%%%%%%%%%%%%%%%%%%%%%%%%%%%%%%

\section*{Data availability}\label{sec:Dataavailability}

All data and analyses associated with this manuscript, including additional data supporting the findings of this study, are provided in the Supplementary Information.

%%%%%%%%%%%%%%%%%%%%%%%%%%%%%%%%%%%%%%%%%%%%%%%%%%%%%%%%%%%%%%%%%%%%%%%%%%%%%%%%%%%%%%%%%%%%%%%%%%%%%%%%%%%%%%%%%%%%%%%%%%%%%%%%%%%%%%%%%%%%%%%%%%%%%%%%%%%

\section*{Acknowledgments}\label{sec:Acknowledgments}

The authors acknowledge funding support from the PEPR-DIADEM AMADEUS Project (ANR-23-PEXD-0003) by the French ANR. S.S. acknowledges funding support from the France Agence Nationale de la recherche (ANR)  associated with the Chair Junior Professor fund (ANR-22-CPJ1-0039-01). S. L. is grateful to the interdisciplinary project IEL “MATPULSE”, part of the French PIA project “Lorraine Université d’Excellence” reference ANR15-IDEX-04-LUE. Calculations were carried out on the national supercomputing resources of the ADASTRA-CINES machines (project No. A0160915123, A0180915123, A0200915123). We also utilized regional supercomputing facilities provided by the Mésocentre de Calcul Intensif Aquitain (MCIA) of the Université de Bordeaux and the Université de Pau et des Pays de l’Adour.

%%%%%%%%%%%%%%%%%%%%%%%%%%%%%%%%%%%%%%%%%%%%%%%%%%%%%%%%%%%%%%%%%%%%%%%%%%%%%%%%%%%%%%%%%%%%%%%%%%%%%%%%%%%%%%%%%%%%%%%%%%%%%%%%%%%%%%%%%%%%%%%%%%%%%%%%%%%

\section*{Author contributions}\label{sec:contributions}

P.U. carried out the computational work, analysed the data, and wrote the manuscript. S.C., S.B., F.P., S.L., and O.M. contributed to data interpretation and discussion. A.B. and S.S. supervised the project, contributed to the conceptual design, and revised the manuscript. All authors approved the final version.

%%%%%%%%%%%%%%%%%%%%%%%%%%%%%%%%%%%%%%%%%%%%%%%%%%%%%%%%%%%%%%%%%%%%%%%%%%%%%%%%%%%%%%%%%%%%%%%%%%%%%%%%%%%%%%%%%%%%%%%%%%%%%%%%%%%%%%%%%%%%%%%%%%%%%%%%%%%

\section*{Competing interests}\label{sec:Competing}

The authors declare no competing interests.

%%%%%%%%%%%%%%%%%%%%%%%%%%%%%%%%%%%%%%%%%%%%%%%%%%%%%%%%%%%%%%%%%%%%%%%%%%%%%%%%%%%%%%%%%%%%%%%%%%%%%%%%%%%%%%%%%%%%%%%%%%%%%%%%%%%%%%%%%%%%%%%%%%%%%%%%%%%

\section*{Additional information}\label{sec:information}

%%%%%%%%%%%%%%%%%%%%%%%%%%%%%%%%%%%%%%%%%%%%%%%%%%%%%%%%%%%%%%%%%%%%%%%%%%%%%%%%%%%%%%%%%%%%%%%%%%%%%%%%%%%%%%%%%%%%%%%%%%%%%%%%%%%%%%%%%%%%%%%%%%%%%%%%%%%

%\bibliographystyle{apsrev4-2}
%\bibliography{references}
%\bibliography{apssamp}
\bibliographystyle{naturemag}
\bibliography{references}

%%%%%%%%%%%%%%%%%%%%%%%%%%%%%%%%%%%%%%%%%%%%%%%%%%%%%%%%%%%%%%%%%%%%%%%%%%%%%%%%%%%%%%%%%%%%%%%%%%%%%%%%%%%%%%%%%%%%%%%%%%%%%%%%%%%%%%%%%%%%%%%%%%%%%%%%%%%

\end{document}